\documentclass[aps,prd,reprint,a4paper,nofootinbib]{revtex4-2}

\usepackage[colorlinks,breaklinks=true,plainpages=false,citecolor=blue,linkcolor=blue,urlcolor=blue,bookmarksopen=true,bookmarksnumbered=false,bookmarksdepth=5]{hyperref}

\usepackage{amssymb}
\usepackage{mathtools}
\usepackage{lmodern}
\usepackage[T1]{fontenc}
\usepackage{bbm}
\DeclareMathAlphabet{\mathbfi}{OML}{cmm}{b}{it}

\let\originalleft\left
\let\originalright\right
\renewcommand{\left}{\mathopen{}\mathclose\bgroup\originalleft}
\renewcommand{\right}{\aftergroup\egroup\originalright}

\makeatletter
\newenvironment{equations}[1][]{\subequations\ifx\relax#1\relax\else\label{#1}\fi\align\ignorespaces}{\endalign\ignorespacesafterend\endsubequations}
\def\@spliteq#1{\begin{equation}\begin{split}#1\end{split}\end{equation}}
\def\splitequation{\collect@body\@spliteq}

\makeatother

\renewcommand{\vec}[1]{{\ifnum9<1#1\mathbf{#1}\else\ifcat\noexpand#1\relax\boldsymbol{#1}\else\mathbfi{#1}\fi\fi}}
\newcommand{\mathe}{\mathrm{e}}
\newcommand{\mathi}{\mathrm{i}}
\newcommand{\total}{\mathop{}\!\mathrm{d}}
\newcommand{\laplace}{\mathop{}\!\bigtriangleup}
\newcommand{\abs}[1]{{\left\lvert{#1}\right\rvert}}
\newcommand{\sgn}{\operatorname{sgn}}
\DeclareMathOperator{\Ein}{Ein}

\newcommand{\1}{\mathbbm{1}}
\newcommand{\eqend}[1]{\,#1}
\newcommand{\bigo}[1]{\mathcal{O}\left({#1}\right)}

\newcommand{\expect}[1]{\left\langle{#1}\right\rangle}
\newcommand{\hankel}[1]{\mathop{}\!\mathrm{H}^{(#1)}}

\allowdisplaybreaks

\begin{document}

\title{Noncommutative geometry from perturbative quantum gravity in de Sitter spacetime}

\author{Markus B. Fr\"ob}
\email{mfroeb@itp.uni-leipzig.de}
\affiliation{Institut f\"ur Theoretische Physik, Universit\"at Leipzig, Br{\"u}derstra{\ss}e 16, 04103 Leipzig, Germany}

\author{William C. C. Lima}
\email{williamcclima@gmail.com}
\affiliation{Institut f\"ur Theoretische Physik, Universit\"at Leipzig, Br{\"u}derstra{\ss}e 16, 04103 Leipzig, Germany}

\author{Albert Much}
\email{much@itp.uni-leipzig.de}
\affiliation{Institut f\"ur Theoretische Physik, Universit\"at Leipzig, Br{\"u}derstra{\ss}e 16, 04103 Leipzig, Germany}

\author{Kyriakos Papadopoulos}
\email[Corresponding author: ]{kyriakos@sci.kuniv.edu.kw}
\affiliation{Department of Mathematics, Kuwait University, Safat 13060, Kuwait}

\begin{abstract}
We show that a noncommutative structure arises naturally from perturbative quantum gravity in a de~Sitter background metric. Our work builds on recent advances in the construction of observables in highly symmetric background spacetimes~[Brunetti et al., \href{https://doi.org/10.1007/JHEP08(2016)032}{JHEP \textbf{08}, 032 (2016)}; Fröb and Lima, \href{https://doi.org/10.1088/1361-6382/aab427}{Class.~Quant.~Grav.~\textbf{35}, 095010 (2018)}], where the dynamical coordinates that are needed in the relational approach were established for such backgrounds to all orders in perturbation theory. We show that these dynamical coordinates that describe events in the perturbed spacetime are naturally noncommuting, and determine their commutator to leading order in the Planck length. Our result generalizes the causal noncommutative structure that was found using the same approach in Minkowski space~[Fröb, Much and Papadopoulos, \href{https://doi.org/10.1103/PhysRevD.107.064041}{Phys.~Rev.~D \textbf{107}, 064041 (2023)}].
\end{abstract}
\maketitle

\section{Introduction}
\label{sec:intro}

The quantization of general relativity (GR) using the well-known approach of perturbative quantum field theory leads to a nonrenormalizable quantum theory of gravity that, in principle, loses all predictive power. However, at scales that are well separated from the fundamental scale (that is, for large distances or low energies compared to the Planck scale), treating it as an effective field theory one can obtain qualitatively and quantitatively meaningful quantum corrections to the classical results~\cite{burgess2003}. This effective approach to quantum gravity, where one quantizes metric fluctuations around a fixed classical background metric, is known as perturbative quantum gravity (pQG). In pQG, the diffeomorphism symmetry of classical GR translates into a gauge transformation for the metric fluctuations, and the resulting gauge theory can be treated using the well-known Becchi--Rouet--Stora--Tyutin formalism~\cite{becchietal1975}.

However, there are some important differences to the usual (Yang--Mills type) gauge theories: as in any diffeomorphism-invariant theory, the identification of observables that describe local measurements is complicated (see, for example, Ref.~\cite{rovelli_2000} for a review). Because diffeomorphisms move points, physical observables cannot be local fields (i.e., defined at a point of the manifold), since by definition observables are gauge invariant. That is, the general covariance of GR implies that the outcome of measurements cannot depend on the arbitrary choice of coordinates, which are changed by diffeomorphisms, and hence physical observables describing these measurements must be diffeomorphism invariant. As an example, consider a scalar field $S(x)$ and a diffeomorphism $x^\mu \to x^\mu - \xi^\mu(x)$. Under this transformation the scalar field changes as $S \to S + \xi^\mu \partial_\mu S$ to first order in $\xi$ and, thus, cannot be an observable unless it is constant.

A possible way around this problem is to consider the framework of relational observables. This approach makes manifest the idea that measurements always concern the state of some dynamical quantity with respect to a different one, for example the gravitational field at the spacetime point where the measurement apparatus sits. In practice, one chooses four dynamical scalar fields that serve as coordinates, in the sense that points are determined by the value that these fields assume there. Observables are then obtained by evaluating the quantity of interest in this dynamical coordinate system, i.e., they are the value of a chosen physical quantity at that point of the manifold where the dynamical scalars that serve as coordinates take on a prescribed value. In other words, what the relational approach shows is that physical events in a gravitational theory cannot be described using coordinates on the background manifold, but instead one must use a fully dynamical coordinate system. This approach goes back a long way in the literature~\cite{komar_pr_1958,bergmann_komar_prl_1960,bergmann_rmp_1961}; see Ref.~\cite{tambornino_sigma_2012} for a recent review. For a sufficiently generic background spacetime, one can, for example, take geometrical scalar fields, such as curvature scalars. In perturbation theory, one then requires that the background value of the chosen scalar fields discriminate points in the background spacetime, and then computes perturbative corrections to both the dynamical coordinates and the relational observables.\footnote{Another way to construct gauge-invariant observables in quantum gravity is to extend the concept of dressed observables in gauge theories~\cite{dirac1955,kibble1968,kulishfaddeev1970,steinmann1984,baganlavellemcmullan2000a,baganlavellemcmullan2000b,mitraratabolesharatchandra2006} to pQG, as proposed in Refs.~\cite{waresaotomeakhoury2013,donnellygiddings2016a,donnellygiddings2016b,giddingsperkins2022}. Interestingly, these dressed observables can be reformulated in the framework of relational observables, see Ref.~\cite{goellerhoehnkirklin2022}.}

However, in the case of perturbative gravity over highly symmetric backgrounds (such as flat Minkowski spacetime or cosmological spacetimes), curvature scalars can clearly not distinguish between all points of the background manifold and thus are ill suited for the task at hand. In other words, relational observables constructed with curvature scalars as dynamical coordinates only measure the average over all points where the curvature scalars are constant and thus cannot describe localized measurements in such background spacetimes. While one can add the required scalar fields by hand (such as the famous Brown--Kucha{\v r} dust~\cite{brownkuchar1995}), this changes the dynamics of the theory~\cite{gieselheroldlisingh2020,giesellisingh2021} and therefore might be undesirable. A suitable method to treat highly symmetric cases without introducing extra fields was first proposed in Ref.~\cite{brunettietal2016} and further developed in Refs.~\cite{froeb2018,froeblima2018,froeblima2022,froeblima2023}. There, the general idea is the construction of the required scalar fields as solutions of scalar field equations in the perturbed spacetime, which are identically satisfied by the background coordinates in the background spacetime. This method is an explicit example of the so-called geometrical clocks~\cite{giesel2008,gieselherzogsingh2018} used in the relational approach and has the advantage that one can easily construct the field-dependent coordinates and corresponding relational observables to arbitrary order in perturbation theory. It has already been employed successfully in pQG for the computation of graviton loop corrections to invariant scalar correlators in Minkowski spacetime~\cite{froeb2018}, of the quantum gravitational backreaction on the Hubble rate during inflation~\cite{froeb2018b,lima2021} and of quantum-gravity corrections to the Newtonian potential of a point particle~\cite{froebreinverch2022}.

Apart from the quantization of gravity as an effective field theory, one can also approach quantum gravity by considering the spacetime to be of quantum nature, i.e., quantizing spacetime itself. This approach is what is commonly referred to as noncommutative geometry, and various different methods are taken under this umbrella. A particular example is the Moyal--Weyl spacetime, where one imposes a quantization condition on the coordinates $x^\mu$ themselves. That is, one promotes the coordinates to self-adjoint operators $\hat{x}^\mu$ and postulates canonical commutation relations~\cite{dfr1995}
\begin{equation}
\label{eq:moyalweyl_commutator}
[ \hat{x}^\mu, \hat{x}^\nu ] = \mathi \Theta^{\mu\nu} \eqend{,}
\end{equation}
where $\Theta$ is a constant skew-symmetric matrix that is assumed to be proportional to the square of the Planck length $\ell_\text{Pl}$. While this example has been studied intensively, there are various issues, among them the UV-IR mixing problem~\cite{minwallavanraamsdonkseiberg2000,matusissusskindtoumbas2000,amelinocameliamandaniciyoshida2004,horvatilakovactrampeticyou2011} (a problem of renormalizability in the formulation of quantum field theories in such spacetimes) and the breaking of Lorentz covariance~\cite{carrollharveykosteleckylaneokamoto2001}.

Even though these two approaches seem very different, recently a connection between them was found in Ref.~\cite{froebmuchpapa22} by three of the authors: the field-dependent dynamical coordinates that are needed to construct gauge-invariant observables in the relational approach define a noncommutative geometry. This connection was realized in the case of pQG over Minkowski spacetime with generalized harmonic coordinates, and the commutator of the dynamical coordinates was computed to leading order in the Planck length. While the resulting commutator has the form~\eqref{eq:moyalweyl_commutator}, there are some important differences: first of all, one does not have a single coordinate operator $\hat{x}^\mu$, but instead associates a dynamical coordinate operator to each physical event. It then turns out that the matrix $\Theta^{\mu\nu}$ is not a spacetime constant, but instead depends on the separation between the events that are described using the dynamical coordinates. Indeed, if they are spacelike separated, $\Theta$ vanishes, while if they are timelike separated $\Theta$ is a constant that depends on which one lies in the future. In this way, the commutator is fully Lorentz invariant and, in particular, vanishes outside the light cone, which in the end follows from the microcausality of the underlying effective field theory that pQG is.

In this work, we generalize the results obtained in Ref.~\cite{froebmuchpapa22} to de~Sitter spacetime. De~Sitter spacetime is important for various reasons. First, it describes to a very good approximation both the inflationary phase of our universe and the current accelerated expansion~\cite{ries_aj_1998,perlmutter_aj_1999,planck_2018}. Second, not all of the de~Sitter manifold is accessible to any single observer: there are both past and future horizons. It thus provides a background spacetime where horizon effects can be studied, with the advantage that it is computationally much easier than a black hole, since de~Sitter is a maximally symmetric spacetime. (However, while the black hole horizon can be defined independent of an observer, the cosmological horizons are observer dependent.) Third, it provides a test bed for different approaches to quantum gravity. In particular, we want to mention the proposition that theories with dynamical gravity contain a reduced number of degrees of freedom when compared to nongravity theories in the same spacetime. Heuristically, this comes from the fact that a black hole is formed when too much mass is concentrated in a region of fixed size, whose entropy is only proportional to the area of the horizon according to the famous Bekenstein--Hawking formula~\cite{bekenstein1972,hawking1975} and not proportional to the volume as for a nongravitational theory. Since the de~Sitter horizon has a finite area, the Bekenstein--Hawking formula also associates to it a finite entropy~\cite{gibbonshawking1977}, whose interpretation is, however, difficult; see~\cite{chandrasekaranetal2023} and references therein for discussions of this issue.\footnote{Assuming a holographic correspondence for de~Sitter space~\cite{strominger2001,alishahihakarchsilversteintong2004,alishahihakarchsilverstein2005}, the horizon entropy can be understood as the entanglement entropy between the right and left dual conformal field theories (CFTs) that appear in the correspondence, or between the past and future dual CFTs~\cite{nguyen2017,narayan2018,dongsilversteintorroba2018,genggrieningerkarch2019,ariasdiazsundell2020,geng2020,geng2021}. We thank Hao Geng for bringing these works to our attention.}

Let us summarize: The aim of this work is to show that one does not need to postulate a noncommutative spacetime such as the Moyal--Weyl one~\eqref{eq:moyalweyl_commutator}. Rather, the noncommutative structure is seen to be a \emph{prediction} of pQG, the well-established effective field theory approach to quantum gravity. The form of the resulting commutator depends, in general, on the background spacetime and the dynamical coordinates, and for a Minkowski background and generalized harmonic coordinates it is a simple and Lorentz-invariant function to leading order in the Planck length~\cite{froebmuchpapa22}. In this article, we want to show that the same holds for a de~Sitter background, which describes the large-scale structure of our universe to a very good approximation. In this way, we will be able to connect the literature that exists for noncommutative spacetimes and their phenomenological implications, see, for example, Ref.~\cite{much2014}, with the pQG approach.

The remainder of the article is structured as follows: In Sec.~\ref{sec:field_dependent_coord}, we give more details on the construction of the required dynamical coordinates, and in Sec.~\ref{sec:field_dependent_coord_dS} we construct them to first order in perturbation theory around a de~Sitter background. In Sec.~\ref{sec:quantum_gravity}, we quantize metric fluctuations around de~Sitter spacetime, and give explicit solutions for the dynamical coordinates as functionals of the metric perturbation. In Sec.~\ref{sec:commutator}, we finally compute the commutator of the dynamical coordinates to leading order in perturbation theory and conclude in Sec.~\ref{sec:discussion}.

\paragraph*{Conventions:} We assume an $n$-dimensional spacetime, use the ``+++'' convention of~\cite{mtw}, and define $\kappa = \sqrt{ 16 \pi } \, \ell_\text{Pl} = \sqrt{ 16 \pi \hbar G_\text{N}/c^3 }$. Greek indices denote spacetime indices, whereas latin indices are purely spatial.

\section{Field-dependent coordinates}
\label{sec:field_dependent_coord}

As discussed in the Introduction, a relational observable corresponds to a dynamical field evaluated at spacetime points that are not fixed, but are determined by the value that other dynamical fields in the system assume there. These other dynamical fields define a (field-dependent) reference frame with respect to which the measurement of a given observable is carried out.

To make these ideas more concrete, let us consider an $n$-dimensional spacetime with metric $g_{\mu\nu}$ and coordinates $x^\mu$. Let us assume that we can find $n$ scalar fields $X^{(\mu)} = X^{(\mu)}[g]$ that are functionals of the spacetime metric.\footnote{We keep the index $\mu$ within parenthesis in $X^{(\mu)}$ to stress that these are a collection of scalar fields.} We further assume that the map $x^\mu \to X^{(\mu)}(x)$ defines a diffeomorphism in our spacetime, such that we can view the dynamical fields $X^{(\mu)}$ as field-dependent coordinates. We can then convert any tensor field into an observable by evaluating it in the field-dependent frame $X^{(\mu)}$.

As a simple example, let us consider a scalar field $S(x)$. The transformation of $S$ to the field-dependent coordinates $X^{(\mu)}(x)$ yields 
\begin{equation}
\label{eq:invariant_scalar}
\mathcal{S}(X) \equiv S[x(X)] \eqend{,}
\end{equation}
where $x^\mu(X)$ denotes the inverse of the map $x^\mu \to X^{(\mu)} = X^{(\mu)}(x)$. Now, consider an arbitrary diffeomorphism $f\colon x^\mu \to (x')^\mu = f^\mu(x)$ with inverse $x^\mu = (f^{-1})^\mu(x')$. Since $X^{(\mu)}$ and $S$ are scalar fields, they both transform under $f$ as
\begin{equations}
X^{(\mu)}(x) &\to (X')^{(\mu)}(x') = X^{(\mu)}(x) = X^{(\mu)}[f^{-1}(x')] \eqend{,} \label{eq:X_transform} \\
S(x) &\to S'(x') = S(x) = S[f^{-1}(x')] \eqend{.} \label{eq:S_transform}
\end{equations}
Note that while the transformed scalar field $S'$ has the same numerical value at the new point $x'$, it differs at the old point $x$: $S'(x) \neq S(x)$. However, if we instead hold the $X^{(\mu)}$ fixed, we obtain
\begin{splitequation}
\label{eq:check_S_invariance}
\mathcal{S}'(X) &= S'[x'(X)] = S[f^{-1}(x'(X))] \\
&= S[x(X)] = \mathcal{S}(X) \eqend{,}
\end{splitequation}
where in the second equality we used the transformation of $S$~\eqref{eq:S_transform} and in the third equality the expression for the inverse diffeomorphism; the expression for $x'(X)$ is obtained by inverting the relation~\eqref{eq:X_transform} for the transformation of the $X^{(\mu)}$. Hence, even though the diffeomorphism $f$ displaces the points $x^\mu$, changing the form of the scalar fields $X^{(\mu)}$ and $S$ with respect to $x^\mu$, the scalar $\mathcal{S}(X)$ is always computed at the same point where the $X^{(\mu)}$ take on a given value.

In what follows, we will be concerned with (quantum) perturbations of the metric over a given spacetime background $g_{\mu\nu}$ covered by coordinates $x^\mu$. Hence, let us write the full metric $\tilde{g}_{\mu\nu}$ as
\begin{equation}
\label{eq:perturbed_metric}
\tilde{g}_{\mu\nu} = g_{\mu\nu} + \kappa h_{\mu\nu} \eqend{,}
\end{equation}
where $h_{\mu\nu}$ is the perturbation. In this case, we can assume that the field-dependent coordinates $X^{(\mu)}$ are chosen such that they agree with the background coordinates $x^\mu$ for $\kappa = 0$, and, in general, can be written as a power series in $\kappa$. We therefore write
\begin{equation}
\label{eq:X_power_series}
X^{(\mu)}(x) = x^\mu + \kappa X^{(\mu)}_{(1)}(x) + \bigo{\kappa^2} \eqend{.}
\end{equation}
Since the background is fixed, we now consider small diffeomorphisms of the form $x^\mu \to x^\mu - \kappa \xi^\mu(x)$, which implies the following change for the metric perturbation:
\begin{equation}
\label{eq:gauge_transf}
\delta_\xi h_{\mu\nu} = \nabla_\mu \xi_\nu + \nabla_\nu \xi_\mu + \bigo{\kappa} \eqend{,}
\end{equation}
where $\nabla_\mu$ is the covariant derivative of the background metric $g_{\mu\nu}$. Moreover, because of the assumption that $X^{(\mu)}$ are scalar fields, we also have that
\begin{equation}
\label{eq:X_transformation}
\delta_\xi X^{(\mu)} = \kappa \xi^\mu + \bigo{\kappa^2} \eqend{.}
\end{equation}
Expanding also the scalar $S$ in a power series in $\kappa$, we can then use Eq.~\eqref{eq:X_transformation} to check explicitly that $\mathcal{S}$ defined in Eq.~\eqref{eq:invariant_scalar} is gauge invariant (at least up to first order).

To construct the field-dependent coordinates $X^{(\mu)}$, we shall follow Refs.~\cite{brunettietal2016,froeb2018,froeblima2018,froeblima2022} and assume that they are solutions of some set of scalar differential equations
\begin{equation}
\label{eq:X_equations}
D^{(\mu)}_{\tilde{g}}(X) = 0 \eqend{,}
\end{equation}
where $D^{(\mu)}_{\tilde{g}}$ are (possibly nonlinear) differential operators involving the spacetime metric $\tilde{g}_{\mu\nu}$. Since they are solutions of differential equations with coefficients involving $\tilde{g}_{\mu\nu}$, we thus expect the dynamical coordinates $X^{(\mu)}$ to be, in general, nonlocal functionals of the metric. Equations~\eqref{eq:X_equations} model the reference frame in which we perform our measurements, which depends on the specific experimental setting we have at hand. We shall furthermore require that Eqs.~\eqref{eq:X_equations} (i) reduce to $D^{(\mu)}_g(x) = 0$ at the background level $\kappa = 0$ and (ii) are causal, i.e., the $X^{(\mu)}(x)$ should only depend on the metric perturbations within the past light cone of the observation point $x$. Condition (i) realizes our assumption that the field-dependent and background frames coincide in the absence of perturbations~\eqref{eq:X_power_series}, while condition (ii) avoids unphysical action-at-a-distance effects coming from arbitrary large spacelike separations~\cite{froeb2018,froeblima2018}.

\section{Field-dependent coordinates for de~Sitter spacetime}
\label{sec:field_dependent_coord_dS}

For the background spacetime, we take the exponentially expanding half of de~Sitter spacetime, the so-called Poincar\'e patch, which is the portion of de~Sitter spacetime most relevant to cosmology. In the Poincar\'e patch, the de~Sitter metric $g_{\mu\nu}$ has the line element
\begin{equation}
\label{eq:dS_metric}
\total s^2 = - \total t^2 + a^2(t) \total \vec{x}^2 \eqend{,}
\end{equation}
where $t \in \mathbb{R}$ is the cosmological time, $a(t) \equiv \mathe^{H t}$ is the scale factor, $H$ is the Hubble constant, and the spatial sections are flat and described using Cartesian coordinates. Clearly, as $H \to 0$ we obtain $a(t) \to 1$ and recover the flat Minkowski spacetime. The Christoffel symbols in these coordinates are given by
\begin{equation}
\label{eq:Christoffel_dS}
\Gamma^\rho_{\mu\nu} = H \left( u^\rho g_{\mu\nu} - u^\rho u_\mu u_\nu - u_\mu \delta^\rho_\nu - u_\nu \delta^\rho_\mu \right) \eqend{,}
\end{equation}
where $u_\mu \equiv - \partial_\mu t = - \delta_\mu^0$. Using the metric~\eqref{eq:dS_metric}, it is easy to verify that the background coordinates $x^\mu = (t,\vec{x})$ satisfy 
\begin{equation}
\nabla^2 x^\mu = - (n-1) H u^\mu \eqend{,}
\end{equation}
where $\nabla^2 \equiv g^{\mu\nu} \nabla_\mu \nabla_\nu$. In analogy to the background coordinates, we now define the field-dependent coordinates $X^{(\mu)}$ on the perturbed spacetime by~\cite{tsamiswoodard2013,lima2021}
\begin{equation}
\label{eq:X_eq}
\tilde{\nabla}^2 X^{(\mu)} = - (n-1) H u^\mu
\end{equation}
and solve this equation perturbatively.

Hence, we take the perturbed metric~\eqref{eq:perturbed_metric}, set
\begin{equations}
X^{(0)}(x) &= t + \kappa X^{(0)}_{(1)}(x) + \bigo{\kappa^2} \eqend{,} \\
X^{(i)}(x) &= x^i + \kappa X^{(i)}_{(1)}(x) + \bigo{\kappa^2} \eqend{,}
\end{equations}
and substitute these expansions into Eq.~\eqref{eq:X_eq}. The equation for the first-order corrections then reads
\begin{splitequation}
\label{eq:X_1_eq}
\nabla^2 X^{(\mu)}_{(1)} &= \nabla_\nu h^{\mu\nu} - \frac{1}{2} \nabla^\mu h - h^{\rho\sigma} \Gamma^\mu_{\rho\sigma} \\
&= g^{\mu\alpha} g^{\nu\beta} \partial_\nu h_{\alpha\beta} - \frac{1}{2} g^{\mu\nu} \partial_\nu h \\
&\quad - (n-3) H u_\nu h^{\mu\nu} + 2 H u^\mu u^\nu u^\rho h_{\nu\rho} \eqend{,}
\end{splitequation}
where $h \equiv g^{\alpha\beta} h_{\alpha\beta}$. Equation~\eqref{eq:X_1_eq} can be solved after initial and boundary conditions have been specified. We will assume that the metric perturbations are either localized (of compact support) or fall off for large spacelike and timelike distances and can then choose the initial conditions
\begin{equation}
\lim_{t \to -\infty} X^{(\mu)}_{(1)}(t,\vec{x}) = \lim_{t \to -\infty} \partial_t X^{(\mu)}_{(1)}(t,\vec{x}) = 0 \eqend{.}
\end{equation}
The solution for Eq.~\eqref{eq:X_1_eq} for a classical metric perturbation $h_{\mu\nu}$ can then be written as
\begin{equation}
\label{eq:X_1_int}
X^{(\mu)}_{(1)}(x) = \int G^\text{ret}(x,x') D^{\mu\alpha\beta} h_{\alpha\beta}(x') \sqrt{-g} \total^n x' \eqend{,}
\end{equation}
where $G^\text{ret}(x,x')$ is the retarded Green's function satisfying
\begin{equation}
\nabla^2 G^\text{ret}(x,x') = \frac{1}{\sqrt{-g}} \delta^n(x-x') \eqend{.}
\end{equation}
Here, we also defined
\begin{equations}
\begin{split}
D^{0\alpha\beta} &\equiv \frac{1}{2} u^\alpha u^\beta \partial_t + \frac{1}{2} a^{-2} \bar\eta^{\alpha\beta} \partial_t - a^{-2} u^{(\alpha} \bar\eta^{\beta)k} \partial_k \\
&\quad+ (n-1) H u^\alpha u^\beta - H a^{-2} \bar\eta^{\alpha\beta} \eqend{,}
\end{split} \\
\begin{split}
D^{i\alpha\beta} &\equiv - a^{-2} \bar\eta^{i(\alpha} u^{\beta)} \partial_t + a^{-4} \bar\eta^{i(\alpha} \bar\eta^{\beta)k} \partial_k \\
&\quad- \frac{1}{2} a^{-2} g^{\alpha\beta} \bar\eta^{ik} \partial_k - (n-3) H a^{-2} u^{(\alpha} \bar\eta^{\beta)i} \eqend{,}
\end{split}
\end{equations}
and $\bar\eta^{\mu\nu} \equiv \eta^{\mu\nu} + u^\mu u^\nu$ is the purely spatial part of the flat metric.

As in Ref.~\cite{froebmuchpapa22}, here we are interested in the commutation relations of the field-dependent coordinates $X^{(\mu)}$ when the metric perturbation is of quantum origin. Since the metric perturbation then has a nonvanishing commutator, by Eq.~\eqref{eq:X_1_int} also the field-dependent coordinates do not commute. We shall employ standard quantum field theory techniques to quantize the metric perturbations in the background de~Sitter spacetime and compute the leading contribution to the commutator.

\section{Quantization of the metric perturbations}
\label{sec:quantum_gravity}

To quantize the metric perturbation, we expand the Einstein--Hilbert action for gravity (including a cosmological constant),
\begin{equation}
\label{eq:fullaction}
S_\text{G} \equiv \kappa^{-2} \int \left( \tilde{R} - 2 \Lambda \right) \sqrt{-\tilde{g}} \total^n x \eqend{,}
\end{equation}
to second order in perturbation theory. In de~Sitter spacetime, the cosmological constant $\Lambda$ is related to the Hubble constant $H$ by $2 \Lambda = (n-1) (n-2) H^2$, and the action~\eqref{eq:fullaction} is simplified if we employ conformally flat coordinates for the background. Hence, we define the conformal time $\eta \in (-\infty,0)$, which is related to the cosmological time as $\eta \equiv - \mathe^{-Ht}/H$. The background metric can then be written in terms of the Minkowski metric $\eta_{\mu\nu}$ as $g_{\mu\nu} = a^2\eta_{\mu\nu}$. After rescaling the metric perturbation as $h_{\mu\nu} = a^2 \hat{h}_{\mu\nu}$, the expansion of the gravitational action~\eqref{eq:fullaction} to second order, $S_\text{G} = S_2 + \bigo{\kappa}$, yields  
\begin{equation}
S_2 = \frac{1}{2} \int a^{n-2} \hat{h}_{\mu\nu} P^{\mu\nu\rho\sigma} \hat{h}_{\rho\sigma} \total^n x \eqend{,}
\end{equation}
where we have defined the differential operator
\begin{splitequation}
\label{eq:pmunurhosigma_def}
P^{\mu\nu\rho\sigma} &\equiv \frac{1}{2} \left[ \eta^{\mu(\rho} \eta^{\sigma)\nu} - \eta^{\mu\nu} \eta^{\rho\sigma} \right] \partial^2 - \partial^{(\mu} \eta^{\nu)(\rho} \partial^{\sigma)} \\
&\quad+ \frac{1}{2} \eta^{\mu\nu} \partial^\rho \partial^\sigma + \frac{1}{2} \eta^{\rho\sigma} \partial^\mu \partial^\nu \\
&\quad+ (n-2) H a \left[ \delta_0^{(\rho} \eta^{\sigma)(\mu} \partial^{\nu)} - \eta^{\mu\nu} \delta^{(\rho}_0 \partial^{\sigma)} \right] \\
&\quad- \frac{n-2}{2} H a \left[ \eta^{\mu(\rho} \eta^{\sigma)\nu} - \eta^{\mu\nu} \eta^{\rho\sigma} \right] \partial_0 \\
&\quad+ \frac{(n-2) (n-1)}{2} H^2 a^2 \eta^{\mu\nu} \delta^\rho_0 \delta^\sigma_0 \eqend{,}
\end{splitequation}
which is symmetric. In this expression, $\partial^2 \equiv \eta^{\mu\nu} \partial_\mu \partial_\nu$ is the flat-space d'Alembertian and indices are raised and lowered with the Minkowski metric $\eta_{\mu\nu}$.

The action $S_2$ is invariant under gauge transformations of the form~\eqref{eq:gauge_transf}, which in conformally flat coordinates and with the above rescaling read
\begin{equation}
\delta_\xi \hat{h}_{\mu\nu} = \partial_\mu \xi_\nu + \partial_\nu \xi_\mu - 2 H a \eta_{\mu\nu} \xi_0 \eqend{.}
\end{equation}
It follows that $P^{\mu\nu\rho\sigma}$ has a nontrivial kernel, i.e., we have $P^{\mu\nu\rho\sigma} \delta_\xi \hat{h}_{\mu\nu} = 0$. Hence, to be able to invert the operator $P^{\mu\nu\rho\sigma}$ and obtain the quantum propagator we need to introduce a gauge-fixing action. Here we follow Ref.~\cite{tsamiswoodard1994} and adopt the gauge-fixing action
\begin{equation}
\label{eq:gauge_fixing_action}
S_\textrm{GF} = - \frac{1}{2} \int H_\mu H^\mu a^{n-2} \total^n x
\end{equation}
with
\begin{equation}
H_\mu \equiv \partial^\nu \hat{h}_{\mu\nu} - \frac{1}{2} \eta^{\rho\sigma} \partial_\mu \hat{h}_{\rho\sigma} - (n-2) H a \hat{h}_{0\mu} \eqend{,}
\end{equation}
which is the de Sitter generalization of the well-known Feynman gauge for the graviton (also called the de Donder gauge in Minkowski spacetime~\cite{capperleibbrandtramonmedrano1973}). We will see that it leads to a similar simplification for the propagator as in flat space~\cite{radkowski1970}. The total action then reads
\begin{equation}
S \equiv S_2 + S_\textrm{GF} = \frac{1}{2} \int a^{n-2} \hat{h}_{\mu\nu} P_\text{GF}^{\mu\nu\rho\sigma} \hat{h}_{\rho\sigma} \total^n x \eqend{,}
\end{equation}
where the differential operator $P_\text{GF}^{\mu\nu\rho\sigma}$ is given by
\begin{splitequation}
P_\text{GF}^{\mu\nu\rho\sigma} &= P^{\mu\nu\rho\sigma} + \partial^{(\mu} \eta^{\nu)(\rho} \partial^{\sigma)} + \frac{1}{4} \eta^{\mu\nu} \eta^{\rho\sigma} \partial^2 - \frac{1}{2} \eta^{\mu\nu} \partial^\rho \partial^\sigma \\
&\quad- \frac{1}{2} \eta^{\rho\sigma} \partial^\mu \partial^\nu - (n-2) H a \delta^{(\rho}_0 \eta^{\sigma)(\mu} \partial^{\nu)} \\
&\quad+ (n-2) H a \eta^{\mu\nu} \delta^{(\rho}_0 \partial^{\sigma)} - \frac{n-2}{4} H a \eta^{\mu\nu} \eta^{\rho\sigma} \partial_0 \\
&\quad+ (n-2) H^2 a^2 \delta_0^{(\mu} \eta^{\nu)(\rho} \delta^{\sigma)}_0 \\
&\quad- \frac{(n-2) (n-1)}{2} H^2 a^2 \eta^{\mu\nu} \delta^\rho_0 \delta^\sigma_0 \\
&= \frac{1}{4} \left[ 2 \eta^{\mu(\rho} \eta^{\sigma)\nu} - \eta^{\mu\nu} \eta^{\rho\sigma} \right] a^2 \nabla^2 \\
&\quad+ (n-2) H^2 a^2 \delta_0^{(\mu} \eta^{\nu)(\rho} \delta^{\sigma)}_0 \eqend{,} \raisetag{8.2em}
\end{splitequation}
where the scalar d'Alembertian $\nabla^2$ in conformally flat coordinates reads
\begin{equation}
a^2 \nabla^2 = \partial^2 - (n-2) H a \partial_\eta \eqend{.}
\end{equation}

The Feynman propagator for the rescaled metric perturbation
\begin{equation}
\label{eq:G_F_tilde}
\hat{G}^\mathrm{F}_{\mu\nu\rho'\sigma'}(x,x') \equiv - \mathi \expect{ \mathcal{T} \, \hat{h}_{\mu\nu}(x) \hat{h}_{\rho'\sigma'}(x') } \eqend{,}
\end{equation}
where $\mathcal{T}$ denotes time ordering and primed indices refer to the coordinate basis at $x'$, satisfies
\begin{equation}
P_\text{GF}^{\alpha\beta\mu\nu} \hat{G}^\mathrm{F}_{\mu\nu\rho'\sigma'}(x,x') = \delta_{(\rho'}^\alpha \delta_{\sigma')}^\beta \frac{\delta^n(x-x')}{a^{n-2}} \eqend{.}
\end{equation}
The advantage of the gauge-fixing action~\eqref{eq:gauge_fixing_action} is that it renders a very simple tensor structure for the graviton propagator. In conformally flat coordinates, the Feynman propagator~\eqref{eq:G_F_tilde} reads~\cite{tsamiswoodard1994,janssenprokopec2008}
\begin{splitequation}
\label{eq:tsamis_woodard_propagator}
\hat{G}^\mathrm{F}_{\mu\nu\rho'\sigma'}(x,x') &= 2 \left[ \bar{\eta}_{\mu(\rho'} \bar{\eta}_{\sigma')\nu} - \frac{1}{n-3} \bar{\eta}_{\mu\nu} \bar{\eta}_{\rho'\sigma'} \right] G^\mathrm{F}_0(x,x') \\
&\quad- 4 \delta_{(\mu}^0 \bar{\eta}_{\nu)(\rho'} \delta_{\sigma')}^0 G^\mathrm{F}_1(x,x') \\
&\quad+ \frac{2}{(n-2) (n-3)} \left[ \eta_{\mu\nu} + (n-2) \delta_\mu^0 \delta_\nu^0 \right] \\
&\qquad\times \left[ \eta_{\rho'\sigma'} + (n-2) \delta_{\rho'}^0 \delta_{\sigma'}^0 \right] G^\mathrm{F}_2(x,x') \eqend{,}
\end{splitequation}
where we recall that $\bar{\eta}_{\mu\nu} = \eta_{\mu\nu} + \delta^0_\mu \delta^0_\nu$ is the purely spatial part of the Minkowski metric. The scalar Feynman propagators $G^\mathrm{F}_s(x,x')$ satisfy
\begin{equation}
\left[ \nabla^2 - s (n-1-s) H^2 \right] G^\mathrm{F}_s(x,x') = \frac{\delta^n(x-x')}{a^n} \eqend{.}
\end{equation} 
They can be written as
\begin{equation}
\label{eq:gs_feynman_def}
G^\mathrm{F}_s(x,x') = \Theta(\eta-\eta') G^+_s(x,x') + \Theta(\eta'-\eta) G^-_s(x,x') \eqend{,}
\end{equation}
where $G^\pm_s(x,x')$ are the Wightman functions, which satisfy $G^-_s(x,x') = G^+_s(x',x)$, and $\Theta$ is the usual Heaviside step function. For later use, we also define the Dyson (or anti-time-ordered) propagator as
\begin{equation}
\label{eq:gs_dyson_def}
G^\mathrm{D}_s(x,x') = \Theta(\eta-\eta') G^-_s(x,x') + \Theta(\eta'-\eta) G^+_s(x,x') \eqend{.}
\end{equation}
Passing to Fourier space according to
\begin{equation}
G^\pm_s(x,x') = \int \tilde{G}^\pm_s(\eta,\eta',\vec{p}) \mathe^{\mathi \vec{p} (\vec{x}-\vec{x'})} \frac{\total^{n-1} \vec{p}}{(2\pi)^{n-1}} \eqend{,}
\end{equation}
in the Euclidean (or Bunch-Davies) vacuum~\cite{chernikovtagirov1968,schomblondspindel1976,bunchdavies1978} we have
\begin{splitequation}
\label{eq:G_plus_s}
\tilde{G}^+_s(\eta,\eta',\vec{p}) &= - \mathi \frac{\pi}{4 H} a^{-\frac{n-1}{2}}(\eta) a^{-\frac{n-1}{2}}(\eta') \\
&\quad\times \hankel{1}_{\frac{n-1}{2} - s}\left( - \abs{\vec{p}} \eta \right) \hankel{2}_{\frac{n-1}{2} - s}\left( - \abs{\vec{p}} \eta' \right) \eqend{,}
\end{splitequation}
where $\hankel{k}_\nu$ is the $k$th Hankel function of order $\nu$~\cite{dlmf}. These expressions simplify for $n = 4$ dimensions, where we obtain
\begin{equations}[eq:G_plus_01]
\tilde{G}^+_0(\eta,\eta',\vec{p}) &= - \mathi \frac{H^2}{2} \frac{\left( 1 + \mathi \abs{\vec{p}} \eta \right) \left( 1 - \mathi \abs{\vec{p}} \eta' \right)}{\abs{\vec{p}}^3} \mathe^{- \mathi \abs{\vec{p}} (\eta-\eta')} \eqend{,} \\
\tilde{G}^+_1(\eta,\eta',\vec{p}) &= \tilde{G}^+_2(\eta,\eta',\vec{p}) = - \mathi \frac{H^2}{2} \frac{\eta \eta'}{\abs{\vec{p}}} \mathe^{- \mathi \abs{\vec{p}} (\eta-\eta')} \eqend{.}
\end{equations}

Since the two-point function $\tilde{G}^\pm_0(\eta,\eta',\vec{p})$ diverges for small $\abs{\vec{p}}$ like $\abs{\vec{p}}^{-3}$, the inverse Fourier transform in three spatial dimensions is not well defined, and the graviton propagator~\eqref{eq:tsamis_woodard_propagator} displays the usual IR divergence of massless fields in the Euclidean vacuum~\cite{fordparker1977,allen1986}. Nevertheless, for the commutator we obtain
\begin{splitequation}
\label{eq:comm_graviton}
\left[ \hat{h}_{\mu\nu}(x), \hat{h}_{\rho\sigma}(x') \right] &= 2 \left[ \bar{\eta}_{\mu(\rho'} \bar{\eta}_{\sigma')\nu} - \bar{\eta}_{\mu\nu} \bar{\eta}_{\rho'\sigma'} \right] \Delta_0(x,x') \1 \\
&\quad+ \Big[ \left( \eta_{\mu\nu} + 2 \delta_\mu^0 \delta_\nu^0 \right) \left( \eta_{\rho'\sigma'} + 2 \delta_{\rho'}^0 \delta_{\sigma'}^0 \right) \\
&\qquad- 4 \delta_{(\mu}^0 \bar{\eta}_{\nu)(\rho'} \delta_{\sigma')}^0 \Big] \Delta_1(x,x') \1 \raisetag{1.6em}
\end{splitequation}
with the state-independent scalar commutator (or Pauli-Jordan) function $\Delta_s$ is well defined also for small $\abs{\vec{p}}$, and convergent in the IR,
\begin{equations}
\tilde{\Delta}_0(\eta,\eta',\vec{p}) &= \mathi H^2 \frac{\eta-\eta'}{\abs{\vec{p}}^2} \cos\left[ \abs{\vec{p}} (\eta-\eta') \right] \\
&\quad- \mathi H^2 \frac{1 + \vec{p}^2 \eta \eta'}{\abs{\vec{p}}^3} \sin\left[ \abs{\vec{p}} (\eta-\eta') \right] \eqend{,} \\
\tilde{\Delta}_1(\eta,\eta',\vec{p}) &= - \mathi H^2 \frac{\eta \eta'}{\abs{\vec{p}}} \sin\left[ \abs{\vec{p}} (\eta-\eta') \right] \eqend{.}
\end{equations}

Using conformal time $\eta$ and the rescaled metric perturbation $\hat{h}_{\mu\nu}$, the explicit expression~\eqref{eq:X_1_int} for the leading quantum correction to the field-dependent coordinates $X^{(\mu)}$ also changes. Performing the change of coordinates and the rescaling, we obtain
\begin{equation}
\label{eq:X_1_int_cc}
X^{(\mu)}_{(1)}(x) = \int G^\text{ret}(x,x') \hat{D}^{\mu\alpha\beta} \hat{h}_{\alpha\beta}(x') a^n(x') \total^n x'
\end{equation}
with
\begin{equations}[eq:hatd_operators]
\begin{split}
\hat{D}^{0\alpha\beta} &= a^{-1} \delta^\alpha_0 \delta^\beta_0 \big[ \partial_\eta + (n-1) H a \big] \\
&\quad- a^{-1} \bar\eta^{k(\alpha} \delta^{\beta)}_0 \partial_k + \frac{1}{2 a} \eta^{\alpha\beta} \partial_\eta \eqend{,}
\end{split} \\
\begin{split}
\hat{D}^{i\alpha\beta} &= - a^{-2} \bar\eta^{i(\alpha} \delta^{\beta)}_0 \big[ \partial_\eta + (n-2) H a \big] \\
&\quad+ a^{-2} \bar\eta^{i(\alpha} \bar\eta^{\beta)k} \partial_k - \frac{1}{2 a^2} \eta^{\alpha\beta} \partial_i \eqend{.}
\end{split}
\end{equations}
Note, however, that the coordinate corrections $X^{(\mu)}_{(1)}$ themselves are not transformed, since they are scalar functionals by construction.

We see from Eq.~\eqref{eq:X_1_int_cc} that the leading quantum correction to the field-dependent coordinates $X^{(\mu)}$ is linear in the metric perturbation $\hat{h}_{\mu\nu}$. Hence, the commutation relation for the quantum fluctuations of the metric~\eqref{eq:comm_graviton} will give rise to nontrivial commutation relations for the coordinates $X^{(\mu)}$ defined by Eq.~\eqref{eq:X_eq}. Nevertheless, since the integral~\eqref{eq:X_1_int_cc} is extended over the full past light cone, additional IR divergences might arise from this integration, and one has to be careful in evaluating the commutator.

\section{Commutator of the generalized harmonic coordinates}
\label{sec:commutator}

To compute the leading contribution to the commutator of the $X^{(\mu)}$, we use the fact that the commutator of a linear field is a state-independent $c$ number, such that we have
\begin{splitequation}
\label{eq:comm_X_1}
&\left[ X^{(\mu)}_{(1)}(x), X^{(\nu)}_{(1)}(x') \right] \\
&\quad= \left[ \expect{ X^{(\mu)}_{(1)}(x) X^{(\nu)}_{(1)}(x') } - \expect{ X^{(\nu)}_{(1)}(x') X^{(\mu)}_{(1)}(x) } \right] \1 \eqend{.} \raisetag{3.8em}
\end{splitequation}
The expectation value on the right-hand side of this equation can be computed in any quantum state. In our case, it is convenient to compute it using the Euclidean vacuum, where the graviton propagator~\eqref{eq:tsamis_woodard_propagator} has a simple expression in Fourier space~\eqref{eq:G_plus_01}. Moreover, at leading order the expectation value on the right-hand side is computed in the free theory, and so we can perform the computation in $n = 4$ dimensions. At higher orders, one would need to consider loop corrections, for which dimensional regularization could be employed with the $n$-dimensional propagators~\eqref{eq:G_plus_s}.

The appropriate formalism to compute true expectation values (rather than scattering matrix elements) is the so-called Schwinger-Keldysh or in-in formalism~\cite{schwinger1961,keldysh1964}. It is particularly useful in calculations in time-dependent spacetimes, as in cosmology, where one is usually interested in expectation values at a given time~\cite{chousuhaolu1985,jordan1986,calzettahu1986,weinberg2005}. In this formalism, one extends the time integration contour: it first goes forward in time from past infinity up to some arbitrary final real time $T$ (which needs to be taken larger than any time one is interested in, and can even be taken to be future infinity), and then backward in time from $T$ to past infinity; it is thus also known as the closed-time path formalism. For practical purposes, it is convenient to split the contour and instead double the number of fields, denoting fields on the forward part of the contour with a ``$+$'' and fields on the backward part of the contour with a ``$-$''. All time integrations then go from $-\infty$ to $+\infty$, and integrations over ``$-$'' fields come with an extra minus sign to compensate for the change of orientation in the backward part of the integration contour. Since contributions from outside the light cone cancel in the integrals, the in-in formalism ensures a causal evolution of observables.

In principle, one can start the integration at a finite time $t_0$ and describe the initial state with a density matrix, which is equivalent to having interaction terms in the action which are localized at $t_0$~\cite{calzettahu1986,cooperetal1994}. Since the commutator~\eqref{eq:comm_X_1} is state independent, we can restrict to the adiabatic vacuum, which can be obtained using the $\mathi \epsilon$ prescription familiar from the in-out formalism~\cite{peskinschroeder} or as the zero-temperature limit of a thermal density matrix~\cite{niegawa1989}. Again, we can choose the simplest route and generalize the $\mathi \epsilon$ prescription. For this, the time integration contour needs to be tilted in the complex plane, with the contours starting and ending at $t_0^\pm = t_0 (1 \mp \mathi \epsilon)$ instead of past infinity. After performing the integral, we take first the limit $t_0 \to - \infty$ with fixed $\epsilon > 0$ and afterward the limit $\epsilon \to 0$.

Instead of time-ordered expectation values, in general, the in-in formalism computes path-ordered expectation values. If all fields lie on the forward part of the contour, i.e., are ``$+$'' fields, these are just time-ordered expectation values, while if all fields are on the backward part, one obtains anti-time-ordered expectation values. Fields on both parts of the contour result in mixed expectation values, with fields on the backward contour always ordered before fields on the forward contour. In particular, the path-ordered two-point function of the (rescaled) metric perturbation reads
\begin{equation}
\hat{G}^{AB}_{\mu\nu\rho\sigma}(x,x') = - \mathi \expect{ \mathcal{P} \, \hat{h}^A_{\mu\nu}(x) \hat{h}^B_{\rho\sigma}(x') }
\end{equation}
with $A,B = \pm$. In line with what was just explained, this propagator can take four values,
\begin{equations}
\begin{split}
\hat{G}^{++}_{\mu\nu\rho\sigma}(x,x') &= - \mathi \expect{ \mathcal{T} \, \hat{h}_{\mu\nu}(x) \hat{h}_{\rho\sigma}(x') } \\
&= \hat{G}^\mathrm{F}_{\mu\nu\rho\sigma}(x,x') \eqend{,}
\end{split} \\
\begin{split}
\hat{G}^{+-}_{\mu\nu\rho\sigma}(x,x') &= - \mathi \expect{ \hat{h}_{\rho\sigma}(x') \hat{h}_{\mu\nu}(x) } \\
&= \hat{G}^-_{\mu\nu\rho\sigma}(x,x') \eqend{,}
\end{split} \\
\begin{split}
\hat{G}^{-+}_{\mu\nu\rho\sigma}(x,x') &= - \mathi \expect{ \hat{h}_{\mu\nu}(x) \hat{h}_{\rho\sigma}(x') } \\
&= \hat{G}^+_{\mu\nu\rho\sigma}(x,x') \eqend{,}
\end{split} \\
\begin{split}
\hat{G}^{--}_{\mu\nu\rho\sigma}(x,x') &= - \mathi \expect{ \overline{\mathcal{T}} \, \hat{h}_{\mu\nu}(x) \hat{h}_{\rho\sigma}(x') } \\
&= \hat{G}^\mathrm{D}_{\mu\nu\rho\sigma}(x,x') \eqend{,}
\end{split}
\end{equations}
encompassing both the time-ordered (Feynman) and anti-time-ordered (Dyson) propagator, as well as the positive and negative frequency Wightman functions. All of these are given by the general formula~\eqref{eq:tsamis_woodard_propagator}, and only the scalar propagators $G_s$ change, with the Feynman and Dyson propagators given by Eqs.~\eqref{eq:gs_feynman_def} and~\eqref{eq:gs_dyson_def}, respectively.

In the quantum theory, also the coordinate corrections $X^{(\mu)}_{(1)}$~\eqref{eq:X_1_int} need to be computed in the in-in formalism, and thus instead of the retarded propagator $G^\text{ret}$ appropriate for the classical theory, we have to use the path-ordered one and sum over both parts of the contour. The path-ordered correlation function of two coordinate corrections thus reads
\begin{splitequation}
\label{eq:xmunu_expect}
&\expect{ \mathcal{P} X^{(\mu)A}_{(1)}(x) X^{(\nu)B}_{(1)}(x') } = \mathi \iint G_0^{AC}(x,y) G_0^{BD}(x',y') \\
&\quad\times \hat{D}_y^{\mu\alpha\beta} \hat{D}_{y'}^{\nu\rho\sigma} \hat{G}^{CD}_{\alpha\beta\rho\sigma}(y,y') \, a^4(y) \total^4 y \, a^4(y') \total^4 y' \eqend{,} \raisetag{1.4em}
\end{splitequation}
where the repeated indices $C,D = \pm$ are summed over, and we recall that integrations over the backward part of the contour come with an extra minus sign. Using the Fourier representation of the relevant propagators~\eqref{eq:tsamis_woodard_propagator} and~\eqref{eq:G_plus_01}, we obtain
\begin{splitequation}
\label{eq:xmunu_expect_fmunu}
&\expect{ \mathcal{P} X^{(\mu)A}_{(1)}(x) X^{(\nu)B}_{(1)}(x') } \\
&\quad= \int F^{\mu\nu}_{AB}(\eta,\eta',\vec{p}) \, \mathe^{\mathi \vec{p} (\vec{x}-\vec{x}')} \frac{\total^3 p}{(2\pi)^3} \eqend{,}
\end{splitequation}
where the components of $F^{\mu\nu}$ are determined in the Appendix and read
\begin{splitequation}
\label{eq:F00}
&F^{00}_{AB}(\eta,\eta',\vec{p}) = \\
& - \iint \frac{\mathe^{- \mathi \abs{\vec{p}} f_{AC}(\eta-\tau)}}{2 \abs{\vec{p}}} \frac{1 + \mathi \abs{\vec{p}} f_{AC}(\eta-\tau) + \vec{p}^2 \eta \tau}{\vec{p}^2} \\
&\qquad\times \frac{\mathe^{- \mathi \abs{\vec{p}} f_{BD}(\eta'-\tau')}}{2 \abs{\vec{p}}} \frac{1 + \mathi \abs{\vec{p}} f_{BD}(\eta'-\tau') + \vec{p}^2 \eta' \tau'}{\vec{p}^2} \\
&\qquad\times \bigg[ 1 + 2 \mathi \abs{\vec{p}} f_{CD}(\tau-\tau') + \mathi \abs{\vec{p}} \tau \tau' \partial_\tau^2 f_{CD}(\tau-\tau') \bigg] \\
&\qquad\times \frac{\mathe^{- \mathi \abs{\vec{p}} f_{CD}(\tau-\tau')}}{2 \abs{\vec{p}}} (\tau \tau')^{-3} \total \tau \total \tau' \eqend{,} \raisetag{1.8em}
\end{splitequation}
\begin{equation}
\label{eq:Fi0}
F^{0i}_{AB}(\eta,\eta',\vec{p}) = F^{i0}_{AB}(\eta,\eta',\vec{p}) = 0 \eqend{,}
\end{equation}
and
\begin{splitequation}
\label{eq:Fij}
&F^{ij}_{AB}(\eta,\eta',\vec{p}) = \\
&\frac{\mathi}{2} \delta_{ij} H^2 \iint \frac{\mathe^{- \mathi \abs{\vec{p}} f_{AC}(\eta-\tau)}}{2 \abs{\vec{p}}} \frac{1 + \mathi \abs{\vec{p}} f_{AC}(\eta-\tau) + \vec{p}^2 \eta \tau}{\vec{p}^2} \\
&\qquad\times \frac{\mathe^{- \mathi \abs{\vec{p}} f_{BD}(\eta'-\tau')}}{2 \abs{\vec{p}}} \frac{1 + \mathi \abs{\vec{p}} f_{BD}(\eta'-\tau') + \vec{p}^2 \eta' \tau'}{\vec{p}^2} \\
&\qquad\times \partial_\tau^2 f_{CD}(\tau-\tau') \, \mathe^{- \mathi \abs{\vec{p}} f_{CD}(\tau-\tau')} (\tau \tau')^{-1} \total \tau \total \tau' \eqend{.}
\end{splitequation}
Here we have introduced the function
\begin{equation}
\label{eq:fcd_def}
f_{CD}(\eta) = \begin{cases} \eta & CD = -+ \\ - \eta & CD = +- \\ \abs{\eta} & CD = ++ \\ - \abs{\eta} & CD = -- \end{cases} \eqend{,}
\end{equation}
and in deriving Eqs.~\eqref{eq:F00}--\eqref{eq:Fij}, we have used that it satisfies
\begin{equations}[eq:fcd_identities]
\left[ \partial_\eta f_{CD}(\eta) \right]^2 &= 1 \eqend{,} \\
\eta \partial_\eta f_{CD}(\eta) &= f_{CD}(\eta)
\end{equations}
for any choice of indices $CD$, as can be easily checked.

Finally, the leading-order contribution to the coordinate commutator~\eqref{eq:comm_X_1} is given by
\begin{splitequation}
\label{eq:nc_x1_expect}
&\left[ X_{(1)}^{(\mu)}(x), X_{(1)}^{(\nu)}(x') \right] \\
&\quad= \int \left[ F^{\mu\nu}_{-+}(\eta,\eta',\vec{p}) - F^{\mu\nu}_{+-}(\eta,\eta',\vec{p}) \right] \mathe^{\mathi \vec{p} (\vec{x}-\vec{x}')} \frac{\total^3 p}{(2\pi)^3} \1 \eqend{.}
\end{splitequation}
As in the case of quantum gravitational perturbations around Minkowski spacetime~\cite{froebmuchpapa22}, the time-space components vanish and the space-space components are proportional to $\delta_{ij}$, showing that also on de~Sitter spacetime only the same coordinate does not commute with itself.

\subsection{Temporal part}
\label{sec:commutator_temporal}

The function $F^{00}_{AB}$~\eqref{eq:F00} is computed in the Appendix, and we obtain
\begin{splitequation}
\label{eq:f00_fourier}
&F^{00}_{+-}(\eta,\eta',\vec{p}) = F^{00}_{-+}(\eta',\eta,\vec{p}) \\
&= \frac{1 + \vec{p}^2 \eta \eta'}{4 \abs{\vec{p}}^3} \mathe^{\mathi \abs{\vec{p}} (\eta+\eta')} \left[ \Ein\left( - 2 \mathi \abs{\vec{p}} \eta' \right) + \gamma + \ln\left( 2 \mathi \abs{\vec{p}} \eta' \right) \right] \\
&+ \frac{1 + \vec{p}^2 \eta \eta'}{4 \abs{\vec{p}}^3} \mathe^{- \mathi \abs{\vec{p}} (\eta+\eta')} \left[ \Ein\left( 2 \mathi \abs{\vec{p}} \eta \right) + \gamma + \ln\left( - 2 \mathi \abs{\vec{p}} \eta \right) \right] \\
&- \frac{\mathe^{\mathi \abs{\vec{p}} (\eta-\eta')}}{8 \abs{\vec{p}}^3} \left[ 1 + 3 \mathi \abs{\vec{p}} (\eta-\eta') - 2 \mathi \abs{\vec{p}} (\eta+\eta') \ln\left( \frac{\eta}{\eta'} \right) \right] \eqend{,}
\end{splitequation}
where the entire function $\Ein$ is given by
\begin{splitequation}
\label{eq:ein_def}
\Ein(z) &\equiv \int_0^1 \frac{\mathe^{z t} - 1}{t} \total t = \sum_{k=1}^\infty \frac{z^k}{k \, k!} \\
&= - \Gamma(0,-z) - \ln(-z) - \gamma
\end{splitequation}
with the incomplete $\Gamma$ function~\cite[Eq.~(8.2.2)]{dlmf}. The expression~\eqref{eq:f00_fourier} is, in fact, not too complicated. It also has the right flat-space limit, as can be seen by going back to cosmological time $t = - H^{-1} \ln( - H \eta )$ and then taking the limit $H \to 0$. This results in
\begin{equation}
\label{eq:f00_fourier_flat}
F^{00}_{+-}(t,t',\vec{p}) = - \frac{1}{4 \abs{\vec{p}}^3} \mathe^{\mathi \abs{\vec{p}} (t-t')} \left[ 1 - \mathi \abs{\vec{p}} (t-t') \right] + \bigo{H} \eqend{,}
\end{equation}
which agrees with the flat-space result~\cite[Eq.~(39)]{froebmuchpapa22}.

It remains to compute the inverse Fourier transform
\begin{splitequation}
\label{eq:f00_fourier_inverse}
F^{00}_{+-}(x,x') &= \int F^{00}_{+-}(\eta,\eta',\vec{p}) \, \mathe^{\mathi \vec{p} (\vec{x}-\vec{x}')} \frac{\total^3 p}{(2\pi)^3} \\
&= \frac{1}{2 \pi^2 r} \lim_{\delta \to 0^+} \int_0^\infty \mathe^{- \delta \abs{\vec{p}}} \abs{\vec{p}} \\
&\qquad\times F^{00}_{+-}(\eta,\eta',\abs{\vec{p}}) \sin\left( \abs{\vec{p}} r \right) \total \abs{\vec{p}}
\end{splitequation}
with $r \equiv \abs{\vec{x}-\vec{x}'}$. Here, we introduced a factor $\mathe^{- \delta \abs{\vec{p}}}$ to ensure convergence for large momenta, and the limit $\delta \to 0$ needs to be taken in the sense of distributions. To compute the inverse Fourier transform, we need the integrals
\begin{subequations}
\begin{align}
&\int_\mu^\infty \frac{1}{p} \mathe^{- a p} \total p = - \gamma - \ln\left( a \mu \right) + \bigo{\mu \ln \mu} \eqend{,} \\
&\int_\mu^\infty \frac{1}{p^2} \mathe^{- a p} \total p = \frac{1}{\mu} - a + a \gamma + a \ln\left( a \mu \right) + \bigo{\mu \ln \mu} \eqend{,} \\
\begin{split}
&\int_0^\infty \mathe^{- a p} \Big[ \Ein\left( \mathi b p \right) + \gamma + \ln\left( - \mathi b p \right) \Big] \total p \\
&\quad= - \frac{1}{a} \ln\left( 1 + \frac{\mathi a}{b} \right) \eqend{,}
\end{split} \\
\begin{split}
&\int_0^\infty \mathe^{- a p} \ln\left( - \mathi b p \right) \Big[ \Ein\left( \mathi b p \right) + \gamma + \ln\left( - \mathi b p \right) \Big] \total p \\
&\quad= \frac{1}{a} \left[ \frac{1}{2} \ln^2\left( 1 + \frac{\mathi a}{b} \right) + \gamma \ln\left( 1 + \frac{\mathi a}{b} \right) - \operatorname{Li}_2\left( - \frac{\mathi a}{b} \right) \right] \eqend{,}
\end{split} \\
\begin{split}
&\int_\mu^\infty \frac{1}{p} \mathe^{- a p} \Big[ \Ein\left( \mathi b p \right) + \gamma + \ln\left( - \mathi b p \right) \Big] \total p \\
&\quad= - \frac{1}{2} \left[ \gamma + \ln\left( - \mathi b \mu \right) \right]^2 - \operatorname{Li}_2\left( - \frac{\mathi a}{b} \right) - \frac{\pi^2}{12} + \bigo{\mu \ln \mu} \eqend{,}
\end{split} \\
\begin{split}
&\int_\mu^\infty \frac{1}{p^2} \mathe^{- a p} \Big[ \Ein\left( \mathi b p \right) + \gamma + \ln\left( - \mathi b p \right) \Big] \total p \\
&\quad= \frac{1}{\mu} \Big[ 1 + \gamma + \ln\left( - \mathi b \mu \right) \Big] + ( a - \mathi b ) \ln \left[ ( a - \mathi b ) \mu \right] \\
&\qquad- a \ln\left( - \mathi b \mu \right) - a + 2 \mathi b - \mathi b \gamma + a \frac{\pi^2}{12} \\
&\qquad+ \frac{a}{2} \left[ \gamma + \ln\left( - \mathi b \mu \right) \right]^2 + a \operatorname{Li}_2\left( - \frac{\mathi a}{b} \right) + \bigo{\mu \ln \mu} \eqend{,} \raisetag{4.6em}
\end{split}
\end{align}
\end{subequations}
valid for $\Re a > 0$, $\Im b = 0$, and $\mu > 0$. These can all be obtained by using the integral form~\eqref{eq:ein_def} for the function $\Ein$, interchanging the integrals over $p$ and $t$, and integration by parts. When computing the integral over $t$, the dilogarithm $\operatorname{Li}_2$ appears, which is defined by~\cite[Eq.~(25.12.2)]{dlmf}
\begin{equation}
\label{eq:dilog_def}
\operatorname{Li}_2(z) = - \int_0^1 \frac{\ln( 1 - z t )}{t} \total t \eqend{,}
\end{equation}
and to simplify the above expressions we also used the identity~\cite[Eq.~(25.12.4)]{dlmf}
\begin{equation}
\operatorname{Li}_2(z) = - \operatorname{Li}_2\left( \frac{1}{z} \right) - \frac{\pi^2}{6} - \frac{1}{2} \ln^2(-z) \eqend{.}
\end{equation}
The introduction of the IR cutoff $\mu$ was necessary because of the before-mentioned IR divergence of massless fields in the Euclidean vacuum, but all terms depending on $\mu$ cancel in the (state-independent) expression for the commutator. Using these integrals, we compute the inverse Fourier transform~\eqref{eq:f00_fourier_inverse}, which results in a lengthy expression that we write down for completeness in App.~\ref{appendix}. It simplifies in the limit $\delta \to 0$, for which we use that in the sense of distributions we have
\begin{equations}
\ln(x \pm \mathi \delta) &\to \ln\abs{x} \pm \mathi \pi \Theta(-x) \eqend{,} \\
\begin{split}
\operatorname{Li}_2(x \pm \mathi \delta) &\to \Theta(1-x) \operatorname{Li}_2(x) + \Theta(x-1) \bigg[ \frac{\pi^2}{6} \\
&\quad- \operatorname{Li}_2(1-x) - \ln(x) \ln(x-1) \pm \mathi \pi \ln x \bigg] \eqend{.}
\end{split}
\end{equations}
Furthermore, many terms cancel in the commutator~\eqref{eq:nc_x1_expect}, since they are symmetric under the exchange of $\eta \leftrightarrow \eta'$. In particular, all terms depending on the IR cutoff $\mu$ completely cancel, and we obtain the IR-finite result
\begin{splitequation}
\label{eq:x00_commutator}
&\left[ X_{(1)}^{(0)}(x), X_{(1)}^{(0)}(x') \right] = \frac{\mathi}{8 \pi} \Theta[ (\eta-\eta')^2 - r^2 ] \sgn(\eta-\eta') \\
&\quad\times \bigg[ \frac{\eta+\eta'}{r} \ln\left( \frac{\eta+\eta'+r}{\eta+\eta'-r} \right) + \ln\left[ \frac{(\eta+\eta')^2 - r^2}{4 \eta \eta'} \right] \\
&\qquad- \frac{3}{2} - \frac{2 \eta \eta'}{r^2 - (\eta+\eta')^2} \bigg] \1 \eqend{.} \raisetag{2em}
\end{splitequation}
This is more complex than the Minkowski result~\cite[Eq.~(47)]{froebmuchpapa22}, but we recover the right flat-space limit by changing back to cosmological time, where for small $H$ we obtain
\begin{splitequation}
\label{eq:x00_commutator_flatlimit}
&\left[ X_1^{(0)}(x), X_1^{(0)}(x') \right] = \frac{\mathi}{8 \pi} \Theta\left[ (t-t')^2 - r^2 \right] \sgn(t-t') \\
&\qquad\times \left[ 1 + \frac{r^2 + 3 (t-t')^2}{24} H^2 + \bigo{H^3} \right] \eqend{.} \raisetag{2em}
\end{splitequation}
We see that as in Minkowski spacetime, the commutator~\eqref{eq:x00_commutator} vanishes outside of the light cone, and its overall sign depends on which of the two points lies in the future of the other. However, in contrast to the flat-space result, it is not constant inside the light cone.

For later use, we can also express the commutator~\eqref{eq:x00_commutator} using the de~Sitter-invariant distance
\begin{equation}
\label{eq:ds_z_def}
Z(x,x') = 1 - \frac{r^2 - (\eta-\eta')^2}{2 \eta \eta'} = \cos\left[ H \mu(x,x') \right] \eqend{,}
\end{equation}
where $\mu$ is the geodesic distance between the two points $x$ and $x'$ and the last equality holds whenever the points are close enough together such that a unique geodesic exists between them.\footnote{If such a geodesic does not exist, then $\mu(x,x') = \pi - \mu(x,\alpha(x'))$ with $\alpha(x')$ the antipodal point of $x'$~\cite{pereznadalrouraverdaguer2010}.} For timelike separation between $x$ and $x'$, we have $Z > 1$, for spacelike separation $Z < 1$, and if the points are lightlike related, we have $Z = 1$. This results in
\begin{splitequation}
\label{eq:x00_commutator_z}
&\left[ X_{(1)}^{(0)}(x), X_{(1)}^{(0)}(x') \right] = \frac{\mathi}{8 \pi} \Theta[ Z(x,x') - 1 ] \sgn(\eta-\eta') \\
&\quad\times \bigg[ \frac{\eta+\eta'}{r} \ln\left[ \frac{r (\eta+\eta') + \eta^2 + (\eta')^2}{\eta \eta' [ 1 + Z(x,x') ]} + \frac{1 - Z(x,x')}{1 + Z(x,x')} \right] \\
&\qquad+ \ln\left[ \frac{1 + Z(x,x')}{2} \right] - \frac{3}{2} + \frac{1}{1 + Z(x,x')} \bigg] \1 \eqend{,} \raisetag{2em}
\end{splitequation}
where now
\begin{equation}
\label{eq:r_in_z}
r = \sqrt{ \eta^2 + (\eta')^2 - 2 \eta \eta' Z(x,x') } \eqend{.}
\end{equation}
That the commutator~\eqref{eq:x00_commutator_z} is not de~Sitter invariant, i.e., that it does not only depend on $Z$, is of course a consequence of the explicit de~Sitter breaking of both the dynamical coordinates~\eqref{eq:X_eq} and the gauge-fixing condition~\eqref{eq:gauge_fixing_action}. However, it becomes invariant for small spatial separation $r \to 0$, where we obtain
\begin{splitequation}
\label{eq:x00_commutator_smallr}
&\left[ X_{(1)}^{(0)}(x), X_{(1)}^{(0)}(x') \right] = \frac{\mathi}{8 \pi} \Theta[ Z(x,x') - 1 ] \sgn(\eta-\eta') \\
&\quad\times \bigg[ \frac{1}{2} + \frac{1}{1 + Z(x,x')} + \ln\left[ \frac{1 + Z(x,x')}{2} \right] + \bigo{r} \bigg] \1 \eqend{.}
\end{splitequation}

\subsection{Spatial part}
\label{sec:commutator_spatial}

The function $F^{ij}_{AB}$~\eqref{eq:Fij} is computed in the Appendix, and we have the quite simple expression
\begin{splitequation}
\label{eq:fij_fourier}
&F^{ij}_{+-}(\eta,\eta',\vec{p}) = \delta_{ij} \frac{H^2}{4 \abs{\vec{p}}^5} \mathe^{\mathi \abs{\vec{p}} (\eta-\eta')} \bigg[ 3 - 3 \mathi \abs{\vec{p}} (\eta-\eta') \\
&\quad- \vec{p}^2 \left( \eta^2 - 3 \eta \eta' + (\eta')^2 \right) - \mathi \abs{\vec{p}}^3 \eta \eta' (\eta-\eta') \bigg] \eqend{.} \raisetag{1.9em}
\end{splitequation}
It also has the right flat-space limit~\cite[Eq.~(39)]{froebmuchpapa22}, which we again can compute by going back to cosmological time $t$ and taking the limit $H \to 0$, resulting in
\begin{equation}
F^{ij}_{+-}(t,t',\vec{p}) = \frac{1}{4 \abs{\vec{p}}^3} \mathe^{\mathi \abs{\vec{p}} (t-t')} \left[ 1 - \mathi \abs{\vec{p}} (t-t') \right] + \bigo{H} \eqend{.}
\end{equation}
We compute the inverse Fourier transform in the same way as in the last subsection and obtain for the commutator in real space
\begin{splitequation}
\label{eq:xij_commutator}
\left[ X_{(1)}^{(i)}(x), X_{(1)}^{(j)}(x') \right] &= \frac{\mathi}{8 \pi} \delta_{ij} \Theta\left[ (\eta-\eta')^2 - r^2 \right] \sgn(\eta-\eta') \\
&\quad\times \frac{H^2}{2} \left[ r^2 - \eta^2 - (\eta')^2 \right] \eqend{.} \raisetag{1.7em}
\end{splitequation}
As for the temporal part~\eqref{eq:x00_commutator_flatlimit}, this commutator agrees with the flat-space result in the limit $H \rightarrow 0$. This is seen changing back to cosmological time $t$, where for small $H$ we obtain
\begin{splitequation}
\label{eq:xij_commutator_flatlimit}
\left[ X_{(1)}^{(i)}(x), X_{(1)}^{(j)}(x') \right] &= - \frac{\mathi}{8 \pi} \delta_{ij} \Theta\left[ (t-t')^2 - r^2 \right] \sgn(t-t') \\
&\quad\times \left[ 1 - (t+t') H + \bigo{H^2} \right] \eqend{,} \raisetag{1.3em}
\end{splitequation}
which agrees with~\cite[Eq.~(47)]{froebmuchpapa22}.

As for the commutator of two time coordinates~\eqref{eq:x00_commutator}, the commutator of two spatial coordinates~\eqref{eq:xij_commutator} vanishes outside of the light cone, its overall sign depends on which of the two points lies in the future of the other, and it is not constant inside the light cone. Expressing it using the de~Sitter invariant $Z$~\eqref{eq:ds_z_def}, we obtain
\begin{splitequation}
\label{eq:xij_commutator_z}
\left[ X_{(1)}^{(i)}(x), X_{(1)}^{(j)}(x') \right] &= - \frac{\mathi}{8 \pi} H^2 \eta \eta' \delta_{ij} \Theta\left[ Z(x,x') - 1 \right] \\
&\quad\times \sgn(\eta-\eta') Z(x,x') \eqend{.} \raisetag{1.3em}
\end{splitequation}
So in contrast to the commutator~\eqref{eq:x00_commutator_z} of the time coordinates, the commutator of spatial coordinates is de~Sitter invariant up to an overall factor.

\section{Discussion}
\label{sec:discussion}

In this work, we have generalized the results obtained in Ref.~\cite{froebmuchpapa22} to de~Sitter spacetime. We quantized perturbations of the metric around a de~Sitter background and computed the dynamical field-dependent coordinates that are needed to describe gauge-invariant observables in the relational framework to linear order. Since the coordinates are (nonlocal) functionals of the quantized metric perturbation $h_{\mu\nu}$, they possess a nonvanishing commutator, and we have computed this commutator to leading order in the Planck length $\ell_\text{Pl}$. Since the coordinates describing the background de~Sitter spacetime do commute among themselves, the leading-order commutator is given by the commutators~\eqref{eq:x00_commutator_z} and \eqref{eq:xij_commutator_z} of the first-order corrections $X_{(1)}^{(\mu)}$. Denoting by $X^{(\mu)} = X^{(\mu)}(p)$ and $Y^{(\mu)} = X^{(\mu)}(q)$ the dynamical coordinates describing two points $p$ and $q$, we thus have (recall that $\kappa^2 = 16 \pi \ell_\text{Pl}^2$)
\begin{splitequation}
\label{eq:commutator}
[ X^{(\mu)}, Y^{(\nu)} ] &= \kappa^2 \left[ X_{(1)}^{(\mu)}(x), X_{(1)}^{(\nu)}(y) \right] + \bigo{\kappa^3} \\
&= 2 \mathi \ell_\text{Pl}^2 \Theta\left[ - (X-Y)^2 \right] \sgn(X^0-Y^0) \\
&\quad\times K^{\mu\nu}(X,Y) \1 + \bigo{\ell_\text{Pl}^3} \raisetag{1.3em}
\end{splitequation}
with
\begin{splitequation}
\label{eq:commutator_temporal}
&K^{00}(x,x') = \ln\left[ \frac{1 + Z(x,x')}{2} \right] - \frac{3}{2} + \frac{1}{1 + Z(x,x')} \\
&\quad+ \frac{\eta+\eta'}{r} \ln\left[ \frac{r (\eta+\eta') + \eta^2 + (\eta')^2}{\eta \eta' [ 1 + Z(x,x') ]} + \frac{1 - Z(x,x')}{1 + Z(x,x')} \right] \eqend{,}
\end{splitequation}
\begin{equation}
K^{0i}(x,y) = 0 \eqend{,}
\end{equation}
and
\begin{equation}
\label{eq:commutator_spatial}
K^{ij}(x,x') = - H^2 \eta \eta' \delta_{ij} Z(x,x') \eqend{,}
\end{equation}
where $Z(x,x')$~\eqref{eq:ds_z_def} is the de~Sitter-invariant distance between $x$ and $x'$, $\eta = - H^{-1} \mathe^{- H t} \in (-\infty,0)$ is the conformal time, and $r$ is the spatial separation, given in terms of $Z$ and $\eta$ by Eq.~\eqref{eq:r_in_z}.

The flat-space limit of the commutator~\eqref{eq:commutator} is obtained by expressing it using the cosmological time $t$ and then taking the limit $H \to 0$. In this limit we have $Z \to 1$ and $K^{\mu\nu} \to \eta^{\mu\nu}$, which can also be read off from Eqs.~\eqref{eq:x00_commutator_flatlimit} and~\eqref{eq:xij_commutator_flatlimit}. The resulting commutator
\begin{splitequation}
\label{eq:commutator_flatspace}
\lim_{H \to 0} [ X^{(\mu)}, Y^{(\nu)} ] &= 2 \mathi \ell_\text{Pl}^2 \Theta\left[ - (X-Y)^2 \right] \sgn(X^0-Y^0) \\
&\quad\times \eta^{\mu\nu} \1 + \bigo{\ell_\text{Pl}^3} \raisetag{1.3em}
\end{splitequation}
agrees with~\cite[Eq.~(48)]{froebmuchpapa22}, which provides a consistency check on our computations. While the spatial part~\eqref{eq:commutator_spatial} is de~Sitter invariant up to an overall factor, the temporal one~\eqref{eq:commutator_temporal} is not. This is a direct consequence of the explicit de~Sitter breaking of both the dynamical coordinates~\eqref{eq:X_eq} and the gauge-fixing condition~\eqref{eq:gauge_fixing_action}. However, de~Sitter invariance of the temporal part~\eqref{eq:commutator_temporal} is recovered for small spatial separation $r \to 0$, and the result is given in Eq.~\eqref{eq:x00_commutator_smallr}.

Apart from the flat-space limit, also the limits of large separation and late times are interesting. Since the commutator~\eqref{eq:commutator} vanishes outside the light cone, only large timelike separations are of interest, which correspond to $Z \to \infty$. To compute the limits, it is useful to express the commutator using the average time $\sigma \equiv (\eta+\eta')/2$ and the difference $\Delta_\eta \equiv \eta-\eta'$. From Eq.~\eqref{eq:x00_commutator}, we then obtain
\begin{splitequation}
&\left[ X_{(1)}^{(0)}(x), X_{(1)}^{(0)}(x') \right] = \frac{\mathi}{8 \pi} \Theta\left( \Delta_\eta^2 - r^2 \right) \sgn \Delta_\eta \\
&\quad\times \bigg[ \frac{2 \sigma}{r} \ln\left( \frac{2 \sigma+r}{2 \sigma-r} \right) + \ln\left[ \frac{4 \sigma^2 - r^2}{4 \sigma^2 - \Delta_\eta^2} \right] \\
&\qquad- \frac{3}{2} - \frac{4 \sigma^2 - \Delta_\eta^2}{2 (r^2 - 4 \sigma^2)} \bigg] \1 \eqend{,}
\end{splitequation}
and from Eq.~\eqref{eq:xij_commutator} we have
\begin{splitequation}
\left[ X_1^{(i)}(x), X_1^{(j)}(x') \right] &= \frac{\mathi}{8 \pi} \delta_{ij} \Theta\left( \Delta_\eta^2 - r^2 \right) \sgn \Delta_\eta \\
&\quad\times \frac{H^2}{4} \left( 2 r^2 - 4 \sigma^2 - \Delta_\eta^2 \right) \1 \eqend{.}
\end{splitequation}
If we keep the spatial separation $r$ constant, the limit of large separation is $\Delta_\eta \to \pm \infty$, in which we have
\begin{equations}
\begin{split}
&\left[ X_{(1)}^{(0)}(x), X_{(1)}^{(0)}(x') \right] \to \frac{\mathi}{8 \pi} \sgn \Delta_\eta \\
&\qquad\times \bigg[ \frac{\Delta_\eta^2}{2 (r^2 - 4 \sigma^2)} - 2 \ln \abs{\Delta_\eta} + \bigo{1} \bigg] \1 \eqend{,}
\end{split} \\
&\left[ X_1^{(i)}(x), X_1^{(j)}(x') \right] \to - \frac{\mathi}{8 \pi} \delta_{ij} \sgn \Delta_\eta \frac{H^2}{4} \left[ \Delta_\eta^2 + \bigo{1} \right] \1 \eqend{.}
\end{equations}
We see that both commutators grow quadratically with the separation, unlike in flat space where the commutator is constant. This indicates a correlation over large distances, which is in fact not too surprising since the dynamical coordinates involve an integral of the metric perturbation over the full past light cone~\eqref{eq:X_1_int}. On the other hand, if we also let the spatial separation grow linearly with the temporal separation, $r = \alpha \Delta_\eta$ with $\abs{\alpha} < 1$, we obtain
\begin{equations}
\begin{split}
&\left[ X_{(1)}^{(0)}(x), X_{(1)}^{(0)}(x') \right] \to \frac{\mathi}{8 \pi} \sgn \Delta_\eta \\
&\quad\times \bigg[ 2 \ln\abs{\alpha} - \frac{3}{2} + \frac{1}{2 \alpha^2} + \bigo{\Delta_\eta^{-1}} \bigg] \1 \eqend{,}
\end{split} \\
\begin{split}
&\left[ X_1^{(i)}(x), X_1^{(j)}(x') \right] \to \frac{\mathi}{8 \pi} \delta_{ij} \sgn \Delta_\eta \\
&\quad\times \frac{H^2}{4} \left[ \left( 2 \alpha^2 - 1 \right) \Delta_\eta^2 + \bigo{1} \right] \1 \eqend{.}
\end{split}
\end{equations}
In this limit, the temporal commutator becomes a constant, while the spatial commutator is still growing. Finally, we consider the late-time limit. Recall that, while the cosmological time $t \in \mathbb{R}$, the conformal time $\eta \in (-\infty,0)$, such that late times correspond to $\sigma \to 0$. In this limit, however, the light cone shrinks to a point, and the commutator vanishes.

As in the flat-space case, the Planck length $\ell_\text{Pl}$ appears naturally in the commutator~\eqref{eq:commutator}, and the commutator is compatible with microcausality, vanishing outside the light cone. The same caveats as there of course also apply to our de~Sitter result: the perturbative effective field theory (EFT) approach is only valid at scales larger than the fundamental scale, the Planck length in our case. On one hand, this makes the commutator~\eqref{eq:commutator} well defined, since to leading order the causal relation between events that enters the right-hand side is the one of the background coordinates, and we do not need to worry how to define causal ordering and a topology for a noncommutative spacetime. On the other hand, we cannot infer strong statements such as the resolution of the singularity in black holes~\cite{nicolinismailagicspalluci2006} from our result, since there the EFT approach breaks down; we could only use our result to make an informed guess, such as in the recent study~\cite{heidarietal2023}.

It would be an interesting but quite tough question to check the validity of our main results for the case in which the background spacetime admits closed timelike curves (CTCs). While Hawking's chronology conjecture~\cite{hawking1992} states that ``\emph{The laws of physics do not allow the appearance of CTCs.}'', taking general relativity in isolation this conjecture does not hold. In particular, there are several solutions of the Einstein equations containing CTCs, such as the van~Stockum dust spacetime and its generalizations~\cite{vanstockum1938,lindsay2016}. However, as advocated by Thorne~\cite{thorne1993}, the combination of general relativity and quantum theory may provide such a mechanism for chronology protection. As a first step and to gain further insight into such a mechanism, one has to study how quantum fields behave in a background spacetime that contains CTCs. However, already quantum mechanics (i.e., nonrelativistic quantum theory) in such a background displays unusual features, and various conditions on the quantum system~\cite{deutsch1991} or even modifications of the postulates of quantum theory have been proposed to deal with these~\cite{hartle1994,hawking1995}. The required conditions and/or modifications can then be generalized to quantum field theory, see, for example, the review~\cite{friedmanhiguchi2006} or Refs.~\cite{pienaarmyersralph2011,tolksdorfverch2018} and references therein.

There are various other questions that remain open. First is the relation to a generalized uncertainty principle (GUP)~\cite{scardigli1999,adlersantiago1999,scardiglicasadio2003,jizbakleinertscardigli2010,tawfikdiab2014,casadioscardigli2014,scardiglilambiasevagenas2017}. Of course it is possible to use the well-known formula
\begin{equation}
\Delta A \, \Delta B \geq \frac{1}{2} \abs{ \expect{ [A, B] } }
\end{equation}
relating the standard deviations $\Delta$ of two Hermitean operators $A$ and $B$ (that is, the uncertainties in the measurement of their values) to the expectation value of their commutator, and thus derive a GUP from our result~\eqref{eq:commutator}. However, while in quantum mechanics one can repeatedly measure the position and momentum of a particle, and from this determine their standard deviations, in quantum gravity it is impossible to measure the coordinates $X^{(\mu)}$ of the \emph{same} event repeatedly. Therefore, the operational meaning of the standard deviation for the coordinates $X^{(\mu)}$ is not clear, and at the moment we do not have anything to say about the relation to possible experiments~\cite{dasvagenas2008,hossenfelder2013,scardiglicasadio2015}. A second question involved higher-order corrections, and going hand in hand with this is the question of how to define a noncommutative spacetime with its topology and causal relations. There are, of course, concrete proposals for noncommutative spacetimes, for example for Riemannian spectral triples~\cite{connes2013}, but the Lorentzian case presents extra challenges~\cite{besnard2009}. As was mentioned in the previous work~\cite{froebmuchpapa22}, already for a classical Lorentzian manifold there can be a mismatch between the topology of the underlying manifold and the causal ordering induced by the Lorentzian metric~\cite{papadopoulos2021}, and we refer the reader to the review~\cite{finstermuchpapadopoulos2021} for an overview of results and open questions. A possible way to define the topology of such a quantum spacetime would be to start on the background spacetime with the path topology of Hawking, King and McCarthy~\cite{hawkingkingmccarthy1976} or one of its refinements~\cite{fullwood1992,low2016,miller2017}, which includes the causal structure of the manifold. In perturbation theory, we have the bijective map~\eqref{eq:X_power_series} from the coordinates of the background spacetime to the dynamical field-dependent coordinates, which then could be used to map the topological structure from the background spacetime to the quantum one. Using the path topology, this would then also give information about the causal structure.

Another question that is worth investigating further concerns the choice of dynamical coordinates. While the concrete choice~\eqref{eq:X_eq} that we made was proposed earlier to construct invariant observables in de~Sitter spacetime~\cite{tsamiswoodard2013,lima2021} and leads to manageable computations, one may wonder if there are other choices that better model a certain physical situation. For example, geodesic light cone coordinates~\cite{prestonpoisson2006,gasperinimarozzinugierveneziano2011,fanizzamarozzimedeirosschiaffino2021,mitsoufanizzagrimmyoo2021,froeblima2022,fanizzamarozzimedeiros2023} model measurements that are made along the observer's past light cone. Also generalized harmonic coordinates $\tilde{\nabla}^2 X^{(\mu)} = 0$ could be a suitable choice, especially since they also appear in other contexts such as matrix models~\cite{steinacker2010}. In fact, the flat-space limit $H \to 0$ of our choice~\eqref{eq:X_eq} are generalized harmonic coordinates, which were used in~\cite{froebmuchpapa22}. It is possible that the commutator of these coordinates then also respects de~Sitter invariance, which is explicitly broken in our case. A related issue is the gauge-fixing dependence of the dynamical coordinates. The diffeomorphism invariance of general relativity translates into a gauge symmetry of the metric perturbations according to Eq.~\eqref{eq:gauge_transf}, which results in the change~\eqref{eq:X_transformation} of the dynamical coordinates. Therefore, as in classical GR, the coordinates themselves do not have an independent physical significance; rather, the invariant observables~\eqref{eq:invariant_scalar} that are constructed using the coordinates are gauge independent~\eqref{eq:check_S_invariance} and physical. Therefore, our result~\eqref{eq:commutator} for the commutator of dynamical coordinates is only part of a physical effect, and further work is needed to construct a suitable observable and connect it to experiments~\cite{dasvagenas2008,hossenfelder2013,scardiglicasadio2015}.

Despite these open questions, our results also clarify some issues that arose in connection with noncommutative spacetimes. First, since the noncommutative structure arises from pQG and is not an extra postulate, it is clear that well-established fundamental physics on large scales is unchanged. Second, the problems of UV-IR mixing~\cite{minwallavanraamsdonkseiberg2000,matusissusskindtoumbas2000,amelinocameliamandaniciyoshida2004,horvatilakovactrampeticyou2011} and the breaking of Lorentz covariance~\cite{carrollharveykosteleckylaneokamoto2001} (in the flat-space limit) are absent, the former because pQG as an effective field theory does not have this issue, and the latter since the matrix $\Theta^{\mu\nu}$ that appears on the right-hand side of the commutation relations~\eqref{eq:moyalweyl_commutator} is not a constant but a function of spacetime, given by Eq.~\eqref{eq:commutator} in de~Sitter or Eq.~\eqref{eq:commutator_flatspace} in Minkowski spacetime. Third, our approach also gives a change of perspective in the interpretation of the commutation relations~\eqref{eq:moyalweyl_commutator}: instead of having a single coordinate operator $\hat{x}^\mu$, in pQG one associates a coordinate operator $X^{(\mu)} = X^{(\mu)}(p)$ to every physical event or point $p$. This change is akin to the change in perspective from quantum mechanics to quantum field theory: instead of quantizing the position and momentum of a single particle, one quantizes canonical coordinates and momenta at each point in spacetime, hence a quantum \emph{field}.

\begin{acknowledgments}
M.B.F.\ acknowledges the support by the Deutsche Forschungsgemeinschaft (DFG, German Research Foundation) --- project no. 396692871 within the Emmy Noether grant CA1850/1-1 and project no. 406116891 within the Research Training Group RTG 2522/1. A.M.\ acknowledges the support by the DFG within the Sonderforschungsbereich (SPP, Priority Program) 2026 ``Geometry at Infinity''.
\end{acknowledgments}

\newpage
\appendix

\section{Details of the computation}
\label{appendix}

To compute the functions $F^{\mu\nu}_{AB}$~\eqref{eq:xmunu_expect_fmunu}, we first need to determine the action of the differential operators $\hat{D}^{\mu\alpha\beta}$~\eqref{eq:hatd_operators} on the propagator~\eqref{eq:tsamis_woodard_propagator}. We obtain
\begin{splitequation}
\label{eq:app_d00}
&a(\eta) a(\eta') \hat{D}^{0\mu\nu}_x \hat{D}^{0\rho\sigma}_{x'} \hat{G}^{CD}_{\mu\nu\rho'\sigma'}(x,x') = - 3 \partial_\eta \partial_{\eta'} G^{CD}_0(x,x') \\
&\qquad+ \big[ 2 \partial_\eta + 3 H a(\eta) \big] \big[ 2 \partial_{\eta'} + 3 H a(\eta') \big] G^{CD}_1(x,x') \\
&\qquad+ \laplace G^{CD}_1(x,x') \eqend{,} \raisetag{1.3em}
\end{splitequation}
\begin{splitequation}
\label{eq:app_d0j}
&a(\eta) a^2(\eta') \hat{D}^{0\mu\nu}_x \hat{D}^{j\rho\sigma}_{x'} \hat{G}^{CD}_{\mu\nu\rho'\sigma'}(x,x') = - \partial_\eta \partial_i G^{CD}_0(x,x') \\
&\qquad- \big[ \partial_{\eta'} + 2 H a(\eta') \big] \partial_i G^{CD}_1(x,x') \eqend{,} \raisetag{1.3em}
\end{splitequation}
\begin{splitequation}
\label{eq:app_di0}
&a^2(\eta) a(\eta') \hat{D}^{i\mu\nu}_x \hat{D}^{0\rho\sigma}_{x'} \hat{G}^{CD}_{\mu\nu\rho'\sigma'}(x,x') = \partial_{\eta'} \partial_i G^{CD}_0(x,x') \\
&\qquad+ \big[ \partial_\eta + 2 H a(\eta) \big] \partial_i G^{CD}_1(x,x') \eqend{,} \raisetag{1.3em}
\end{splitequation}
and
\begin{splitequation}
\label{eq:app_dij}
&a^2(\eta) a^2(\eta') \hat{D}^{i\mu\nu}_x \hat{D}^{j\rho\sigma}_{x'} \hat{G}^{CD}_{\mu\nu\rho'\sigma'}(x,x') = - \delta^{ij} \laplace G^{CD}_0(x,x') \\
&\qquad- \delta^{ij} \big[ \partial_\eta + 2 H a(\eta) \big] \big[ \partial_{\eta'} + 2 H a(\eta') \big] G^{CD}_1(x,x') \eqend{,}
\end{splitequation}
where we used that in $n = 4$ dimensions the propagators $G_1$ and $G_2$ are equal~\eqref{eq:G_plus_01}. We then pass to Fourier space and insert the explicit expressions of the propagators. For the positive frequency Wightman functions $G^{-+}$, these are exactly the expressions given in Eq.~\eqref{eq:G_plus_01}, while for the Feynman $G^{++}$ and Dyson $G^{--}$ propagators we use the relations~\eqref{eq:gs_feynman_def} and~\eqref{eq:gs_dyson_def}. This results in
\begin{equations}
\begin{split}
\tilde{G}^{CD}_0(\eta,\eta',\vec{p}) &= - \mathi \frac{H^2}{2} \frac{1 + \mathi \abs{\vec{p}} f_{CD}(\eta-\eta') + \vec{p}^2 \eta \eta'}{\abs{\vec{p}}^3} \\
&\quad\times \mathe^{- \mathi \abs{\vec{p}} f_{CD}(\eta-\eta')} \eqend{,} \raisetag{1.3em}
\end{split} \\
\tilde{G}^{CD}_1(\eta,\eta',\vec{p}) &= - \mathi \frac{H^2}{2} \frac{\eta \eta'}{\abs{\vec{p}}} \mathe^{- \mathi \abs{\vec{p}} f_{CD}(\eta-\eta')} \eqend{,}
\end{equations}
where the function $f_{CD}$ is defined in Eq.~\eqref{eq:fcd_def}. We then obtain the expressions~\eqref{eq:F00} for $F^{00}_{AB}$, \eqref{eq:Fi0} for $F^{0j}_{AB}$ and $F^{i0}_{AB}$, and~\eqref{eq:Fij} for $F^{ij}_{AB}$, using also the identities~\eqref{eq:fcd_identities}.

To compute the integrals over $\tau$ and $\tau'$ in Eqs.~\eqref{eq:F00} and~\eqref{eq:Fij}, we sum over the indices $C,D = \pm$ and integrate from $\eta_0^\pm$ to a final time $T$ as explained in Sec.~\ref{sec:commutator}, taking care to add minus signs for the ``$-$'' fields to compensate for the change of orientation in the backward part of the integration contour. The resulting integrals contain terms involving, e.g., $\abs{\eta-\tau}$, which we treat by splitting the $\tau$ integral at $\tau = \eta$. One then sees explicitly that any dependence on the final time $T$ cancels, since the in-in formalism ensures a causal evolution. For the second derivative of $f_{CD}$, we use that
\begin{equations}
\partial_\tau^2 f_{-+}(\tau-\tau') &= \partial_\tau^2 f_{+-}(\tau-\tau') = 0 \eqend{,} \\
\partial_\tau^2 f_{++}(\tau-\tau') &= - \partial_\tau^2 f_{--}(\tau-\tau') = 2 \delta(\tau-\tau') \eqend{,}
\end{equations}
and altogether we obtain
\begin{widetext}
\begin{splitequation}
F^{00}_{+-}(\eta,\eta',\vec{p}) &= \frac{1}{8 \abs{\vec{p}}^7} ( 1 + \mathi \abs{\vec{p}} \eta ) \left[ - ( 1 + \mathi \abs{\vec{p}} \eta' ) \, \mathe^{- \mathi \abs{\vec{p}} (\eta+\eta')} G^{(1)}_{+-}(\eta,\eta',\vec{p}) + ( 1 - \mathi \abs{\vec{p}} \eta' ) \, \mathe^{- \mathi \abs{\vec{p}} (\eta-\eta')} G^{(2)}_{+-}(\eta,\eta',\vec{p}) \right] \\
&\quad+ \frac{1}{8 \abs{\vec{p}}^7} ( 1 - \mathi \abs{\vec{p}} \eta ) \left[ ( 1 + \mathi \abs{\vec{p}} \eta' ) \, \mathe^{\mathi \abs{\vec{p}} (\eta-\eta')} G^{(3)}_{+-}(\eta,\eta',\vec{p}) - ( 1 - \mathi \abs{\vec{p}} \eta' ) \, \mathe^{\mathi \abs{\vec{p}} (\eta+\eta')} G^{(4)}_{+-}(\eta,\eta',\vec{p}) \right] \\
\end{splitequation}
with
\begin{splitequation}
G^{(1)}_{+-}(\eta,\eta',\vec{p}) &= 2 \mathi \abs{\vec{p}} \int_{\eta_0^+}^\eta \mathe^{2 \mathi \abs{\vec{p}} \tau} ( 1 - \mathi \abs{\vec{p}} \tau )^2 \tau^{-4} \total \tau \\
&\quad+ \int_{\eta_0^+}^\eta \int_{\eta_0^+}^\tau \mathe^{2 \mathi \abs{\vec{p}} \tau'} ( 1 - \mathi \abs{\vec{p}} \tau ) ( 1 - \mathi \abs{\vec{p}} \tau' ) [ 1 + 2 \mathi \abs{\vec{p}} (\tau-\tau') ] (\tau \tau')^{-3} \total \tau' \total \tau \\
&\quad+ \int_{\eta_0^+}^\eta \int_\tau^{\eta'} \mathe^{2 \mathi \abs{\vec{p}} \tau} ( 1 - \mathi \abs{\vec{p}} \tau ) ( 1 - \mathi \abs{\vec{p}} \tau' ) [ 1 - 2 \mathi \abs{\vec{p}} (\tau-\tau') ] (\tau \tau')^{-3} \total \tau' \total \tau \eqend{,}
\end{splitequation}
\begin{equation}
G^{(2)}_{+-}(\eta,\eta',\vec{p}) = \int_{\eta_0^+}^\eta \int_{\eta_0^-}^{\eta'} \mathe^{2 \mathi \abs{\vec{p}} (\tau-\tau')} ( 1 - \mathi \abs{\vec{p}} \tau ) ( 1 + \mathi \abs{\vec{p}} \tau' ) [ 1 - 2 \mathi \abs{\vec{p}} (\tau-\tau') ] (\tau \tau')^{-3} \total \tau' \total \tau \eqend{,}
\end{equation}
\begin{splitequation}
G^{(3)}_{+-}(\eta,\eta',\vec{p}) &= 2 \mathi \abs{\vec{p}} \int_{\eta'}^\eta ( 1 + \vec{p}^2 \tau^2 ) \tau^{-4} \total \tau \\
&\quad+ \int_{\eta_0^-}^\eta \int_{\eta_0^+}^\tau \mathe^{- 2 \mathi \abs{\vec{p}} (\tau-\tau')} ( 1 + \mathi \abs{\vec{p}} \tau ) ( 1 - \mathi \abs{\vec{p}} \tau' ) [ 1 + 2 \mathi \abs{\vec{p}} (\tau-\tau') ] (\tau \tau')^{-3} \total \tau' \total \tau \\
&\quad+ \int_{\eta_0^-}^{\eta'} \int_\tau^{\eta'} \mathe^{- 2 \mathi \abs{\vec{p}} (\tau-\tau')} ( 1 + \mathi \abs{\vec{p}} \tau ) ( 1 - \mathi \abs{\vec{p}} \tau' ) [ 1 + 2 \mathi \abs{\vec{p}} (\tau-\tau') ] (\tau \tau')^{-3} \total \tau' \total \tau \\
&\quad+ \int_\eta^{\eta'} \int_{\eta'}^\tau ( 1 + \mathi \abs{\vec{p}} \tau ) ( 1 - \mathi \abs{\vec{p}} \tau' ) [ 1 - 2 \mathi \abs{\vec{p}} (\tau-\tau') ] (\tau \tau')^{-3} \total \tau' \total \tau \eqend{,}
\end{splitequation}
and
\begin{splitequation}
G^{(4)}_{+-}(\eta,\eta',\vec{p}) &= - 2 \mathi \abs{\vec{p}} \int_{\eta_0^-}^{\eta'} \mathe^{- 2 \mathi \abs{\vec{p}} \tau} ( 1 + \mathi \abs{\vec{p}} \tau )^2 \tau^{-4} \total \tau \\
&\quad+ \int_{\eta_0^-}^{\eta'} \int_\tau^\eta \mathe^{- 2 \mathi \abs{\vec{p}} \tau} ( 1 + \mathi \abs{\vec{p}} \tau ) ( 1 + \mathi \abs{\vec{p}} \tau' ) [ 1 + 2 \mathi \abs{\vec{p}} (\tau-\tau') ] (\tau \tau')^{-3} \total \tau' \total \tau \\
&\quad+ \int_{\eta_0^-}^{\eta'} \int_{\eta_0^-}^\tau \mathe^{- 2 \mathi \abs{\vec{p}} \tau'} ( 1 + \mathi \abs{\vec{p}} \tau ) ( 1 + \mathi \abs{\vec{p}} \tau' ) [ 1 - 2 \mathi \abs{\vec{p}} (\tau-\tau') ] (\tau \tau')^{-3} \total \tau' \total \tau
\end{splitequation}
for the temporal part, and
\begin{splitequation}
F^{ij}_{+-}(\eta,\eta',\vec{p}) &= \delta_{ij} \frac{\mathi H^2}{4 \abs{\vec{p}}^6} (1 + \mathi \abs{\vec{p}} \eta) (1 + \mathi \abs{\vec{p}} \eta') \int_{t_0^+}^\eta \mathe^{- \mathi \abs{\vec{p}} (\eta+\eta'-2\tau)} \frac{(1 - \mathi \abs{\vec{p}} \tau)^2}{\tau^2} \total \tau \\
&- \delta_{ij} \frac{\mathi H^2}{4 \abs{\vec{p}}^6} (1 - \mathi \abs{\vec{p}} \eta)  (1 - \mathi \abs{\vec{p}} \eta') \int_{t_0^-}^{\eta'} \mathe^{\mathi \abs{\vec{p}} (\eta+\eta'-2\tau)} \frac{(1 + \mathi \abs{\vec{p}} \tau)^2}{\tau^2} \total \tau \\
&+ \delta_{ij} \frac{\mathi H^2}{4 \abs{\vec{p}}^6} (1 - \mathi \abs{\vec{p}} \eta) (1 + \mathi \abs{\vec{p}} \eta') \, \mathe^{\mathi \abs{\vec{p}} (\eta-\eta')} \int_\eta^{\eta'} \frac{1 + \vec{p}^2 \tau^2}{\tau^2} \total \tau
\end{splitequation}
for the spatial part.
\end{widetext}
The integrals can then be done straightforwardly, using the special function $\Ein$ defined in Eq.~\eqref{eq:ein_def}. Finally we set $\eta_0^\pm = \eta_0 (1 \mp \mathi \epsilon)$ and take first the limit $\eta_0 \to -\infty$ with fixed $\epsilon > 0$ and afterward the limit $\epsilon \to 0$. For this limit, we need the asymptotic expansion of $\Ein$, which can be obtained from the one of the incomplete $\Gamma$ function~\cite[Eq.~(8.11.2)]{dlmf} and reads (for $\Re \alpha < 0$ and as $r \to \infty$)
\begin{splitequation}
\Ein(\alpha r + \beta) &\sim - \gamma - \ln\left[ - (\alpha r + \beta) \right] \\
&\quad+ \mathe^{\alpha r + \beta} \left[ \frac{1}{\alpha r} + \bigo{r^{-2}} \right] \eqend{.}
\end{splitequation}
This gives the results~\eqref{eq:f00_fourier} for $F^{00}_{+-}$ and~\eqref{eq:fij_fourier} for $F^{ij}_{+-}$, and the analogous computation for $F^{\mu\nu}_{-+}$ shows that the same result is obtained with $\eta$ and $\eta'$ exchanged.

Finally, we need to compute the inverse Fourier transform~\eqref{eq:f00_fourier_inverse} as explained in Sec.~\ref{sec:commutator_temporal}. The full expression, including the IR cutoff $\mu$ and the UV convergence factor depending on $\delta$, reads
\begin{widetext}
\begin{splitequation}
F^{00}_{+-}(x,x') &= \frac{1}{16 \pi^2} \left[ 3 + \frac{\pi^2}{6} + \gamma + 2 \ln\left( 4 \eta \eta' \right) - \left[ \gamma + \ln \mu + \ln(-2\eta') \right]^2 - \left[ \gamma + \ln \mu + \ln(-2\eta) \right]^2 - \mathi \pi \ln\left( \frac{\eta}{\eta'} \right) + \ln \mu \right] \\
&- \frac{\mathi}{16 \pi^2 r} \bigg[ \frac{1}{2} \left[ - 2 \mathi (\eta+\eta') \ln\left( \frac{\eta}{\eta'} \right) + 3 \delta \right] \Big[ \ln\left[ \delta - \mathi (\eta-\eta'+r) \right] - \ln\left[ \delta - \mathi (\eta-\eta'-r) \right] \Big] \\
&\hspace{4em}- \frac{3}{2} \mathi r \Big[ \ln\left[ \delta - \mathi (\eta-\eta'+r) \right] + \ln\left[ \delta - \mathi (\eta-\eta'-r) \right] \Big] \\
&\hspace{4em}+ [ \delta - \mathi (\eta+\eta') - \mathi r ] \operatorname{Li}_2\left( \frac{\mathi \delta + (\eta+\eta'+r)}{2 \eta'} \right) - [ \delta - \mathi (\eta+\eta') + \mathi r ] \operatorname{Li}_2\left( \frac{\mathi \delta + (\eta+\eta'-r)}{2 \eta'} \right) \\
&\hspace{4em}- [ \delta + \mathi (\eta+\eta') + \mathi r ] \operatorname{Li}_2\left( \frac{- \mathi \delta + (\eta+\eta'+r)}{2 \eta} \right) + [ \delta + \mathi (\eta+\eta') - \mathi r ] \operatorname{Li}_2\left( \frac{- \mathi \delta + (\eta+\eta'-r)}{2 \eta} \right) \\
&\hspace{4em}- \frac{\eta \eta'}{[ \delta - \mathi (\eta+\eta'+r) ]} \ln\left( 1 - \frac{\mathi \delta + (\eta+\eta'+r)}{2 \eta'} \right) + \frac{\eta \eta'}{\delta - \mathi (\eta+\eta'-r)} \ln\left( 1 - \frac{\mathi \delta + (\eta+\eta'-r)}{2 \eta'} \right) \\
&\hspace{4em}+ \frac{\eta \eta'}{[ \delta + \mathi (\eta+\eta'+r) ]} \ln\left( 1 + \frac{\mathi \delta - (\eta+\eta'+r)}{2 \eta} \right) - \frac{\eta \eta'}{\delta + \mathi (\eta+\eta'-r)} \ln\left( 1 + \frac{\mathi \delta - (\eta+\eta'-r)}{2 \eta} \right) \bigg] \eqend{,}
\end{splitequation}
with the dilogarithm $\operatorname{Li}_2$ defined in Eq.~\eqref{eq:dilog_def}.
\end{widetext}

\bibliography{literature}
\end{document}